\documentclass[onecolumn]{article}
\usepackage[english]{babel}

\usepackage[utf8]{inputenc}
\usepackage[pdftex]{graphicx}
\usepackage{authblk}
\usepackage{dcolumn}
\usepackage{amsmath}
\usepackage{longtable}
\usepackage{tabularx}
\usepackage{setspace}
\usepackage{multirow}
\usepackage{mathtools}
\usepackage{tikz}
\usepackage{lscape}
\usepackage{amssymb}
\usepackage{url}
\usepackage{arydshln}

\usepackage{csquotes}
\MakeOuterQuote{"}
\usepackage{lscape}

\usepackage[font=small]{caption}

\RequirePackage[twoside,letterpaper,includeheadfoot,layoutsize={8.125in,10.875in},layouthoffset=0.1875in,layoutvoffset=0.0625in,left=38.5pt,right=43pt,top=43pt,bottom=32pt,headheight=0pt,headsep=10pt,footskip=25pt]{geometry}
\title{Strategic reciprocity improves academic performance in public elementary school children}

\author[1,2,3,*]{\small Cristian Candia}

\author[3]{ V\'ictor Landaeta-Torres}

\author[4,5,6]{C\'esar A. Hidalgo}

\author[3,*]{Carlos Rodriguez-Sickert}

\affil[1]{Kellogg School of Management, Northwestern University, Evanston, IL 60208}
\affil[2]{Northwestern Institute on Complex Systems (NICO), Northwestern University, Evanston, IL 60208}

\affil[3]{Centro de Investigación en Complejidad Social (CICS), Facultad de Gobierno, Universidad del Desarrollo, Las Condes, Chile.}

\affil[4]{ANITI Chair, University of Toulouse, Toulouse, France}

\affil[5]{Alliance Manchester Business School, University of Manchester, Manchester, UK}

\affil[6]{School of Engineering and Applied Sciences, Harvard University}

\affil[*]{Corresponding Authors: cristian.candia@kellogg.northwestern.edu, carlosrodriguez@udd.cl}

\date{\today}    
\begin{document}

\maketitle

\begin{abstract}

Social networks are pivotal for learning. Yet, we still lack a full understanding of the mechanisms connecting networks with learning outcomes. Here, we present the results of a large scale study (946 elementary school children from 45 different classrooms) designed to understand the social strategies used by elementary school children. We mapped the social networks of students using both, a non-anonymous version of a prisoner’s dilemma and a survey of nominated friendships, and compared the strategies played by students with their GPAs. We found that higher GPA students invest more strategically in their relationships, cooperating more generously with friends and less generously with non-friends than lower GPA students.  Our findings suggest that the higher selectivity of social capital investments by high performing students, may be one of the mechanisms helping them reap the learning benefits of their social networks.

\end{abstract}

\section*{Introduction}

In recent decades, several studies have shown a significant correlation between a student's position in a social network and their academic performance \cite{Baldwin1997, Caprara2000, Mouw2006, 8_Bruun, Gasevic2013, Blansky2013, Ivaniushina2018, Stadtfeld2019, Kassarnig2018}. Among college students, academic performance has been shown to correlate positively with social capital in masters students taking online degrees \cite{Gasevic2013}, and with the flow of online and offline communication among undergraduates in Denmark \cite{Kassarnig2018}. Yet, we still lack a complete understanding of the connection between social networks and learning outcomes, specially among young children. 

Here, we study the link between a child's position in their classroom network and their academic performance by mapping the networks of 946 children aged 9 to 11 from 45 classrooms in 14 Chilean public schools. We mapped networks using both, an experimental game theory setup based on a prisoner's dilemma, and surveys where students nominate their friends and the classmates they like or dislike \cite{Ryan2001}. In the game theoretical setup, children use a tablet interface to play a non-anonymous and simultaneous dyadic game with each of their classmates. In the game, each student has to choose how many tokens to send (from 1 to 10) to other student. Since sent tokens double, pairs of students who shared all of their tokens receive twenty tokens at the end of the round. Yet, a student who sent all of their tokens and received nothing in return gets zero tokens, while the non-cooperating student in that situation gets 30 (their own 10 plus 20 from the cooperating student). When neither student cooperates, they keep their initial ten tokens. By using a non-anonymous game, we map the cooperative network among students.

The nomination survey allowed us to identify four types of friendships: i) Mutual friendship, where both the sender and the receiver declared friendship; ii) Ego-asymmetrical friendship, where only the sender student declares friendship; iii) Alter-asymmetrical friendship, where only the receiving student declares friendship; and iv) Non-friendship, when no friendship is declared.

For student performance we used grade point averages (GPAs). GPAs are a comparable measure within a classroom, since at this grade level all students take all classes together with the same teachers. 

We use this data to study how the strategies played by low and high GPA students vary with friendship relationships. We find that students are more cooperative when a friendship signaling exists. Moreover, we find that students are good at predicting when others declares them as friends (alter-asymmetrical friendship), sending more tokens to them than to non-friends, but less than to mutual friends.

We find that high GPA students tend to be more strategic in their cooperative "investments", cooperating comparatively less with non-friends than low GPA students. Interestingly, computing the reciprocated rate of cooperation (reciprocated cooperation over sent cooperation), we find that the benefits of social capital investment in alter-asymmetrical friendship are the same as investment in mutual friendship. In contrast, social capital investment in students with considerably higher GPA is not efficient as it shows the lowest cooperative reciprocation, comparable with non-friendship relationships. 

These results suggest that the selectivity of high GPA students may help them reap the learning benefits of their social networks.

\subsection*{Networks and Learning}

Long-term returns to education depend mostly on early learning outcomes \cite{Claessens2013, Dickens2006, Restuccia2004}, highlighting the need to understand how social and environmental factors contribute to early learning. The literature on learning has identified several factors that contribute to learning at both the individual and group levels.

At the individual level, academic achievement has been shown to improve with parental education \cite{Ahmad2013} and the involvement of parents in school activities \cite{Marcon1999}. Student time management \cite{Macan1990}, sleep quality and duration \cite{Curcio2006}, physical activity \cite{Singh2012} and, in low-income families, internet use \cite{Jackson2006}, also contribute positively to learning.

At the group-level, peer effects are known to be pivotal for social learning \cite{Henrich2015book, Henrich2001, pentland2015social, johnson1987learning, roger1994overview}, playing a key role in academic outcomes \cite{Kassarnig2018, Blansky2013, Sacerdote2011, Davies2018, 10_Biancani}. Indeed, the way in which people capture the effects of their social relationships within a particular social environment, largely depends on the social structure of their cooperative relationships \cite{Calvo-Armengol2009}. For instance, at the dyadic level, social learning emerges as a natural form of pedagogy, where cognitive mechanisms enable the transmission of cultural knowledge through imitation and communication  \cite{csibra2011natural}. What distinguishes this natural pedagogy from other types of social learning, e.g., prestige-biased social learning \cite{henrich2001evolution, 11_McFarland}, is that it not only requires the disposition to learn from the "student-role subject," but the willingness of the "teacher-role subject" to share his/her knowledge \cite{csibra2011natural}. Thus, a pedagogic act between peers \cite{Rohrbeck2003}, qualifies as an act of cooperation in a standard social dilemma from a game-theoretical perspective. 

Capturing cooperative dispositions between individuals is challenging, and often requires multidimensional instruments that enable us to uncover social relationships. Efforts to capture social networks date back to Jacob Moreno's sociograms \cite{Moreno1934}. Under Moreno's approach, the social network is obtained through surveys that ask students to state both, who they like or don't like to spend time with, and who their friends are \cite{Coie1990, Mouw2006, Neal2007, McCormick2015, Ivaniushina2018}. But survey-based instruments have issues, such as the social desirability bias \cite{van2008faking} (over-reporting of socially desirable behavior); cognitive barriers (difficult to establish that subjects fully understand the questions); and lack of engagement associated with the implementation of the instruments (length or unfriendliness of instruments generate poor answers). An extra challenge is to study populations as young as primary school children. Traditionally, relational studies in older populations are conducted through surveys. However, for primary school students, survey-based social network mapping may not be sufficient because of the exacerbation of the aforementioned biases \cite{de2011improving}.

By using a cooperative video game, our methodological approach facilitates the mapping of the cooperative relationships. The advantage of using game theory to map the social network is twofold: first, it allows us to specifically capture the nature of cooperative relationships among students; and second, the interactive nature of the game in which different actions lead to different payoffs partially mitigates the aforementioned issues of survey-based instruments \cite{camerer1999effects}. In the next section, we explain in more detail the game-theoretical experiment, the video game, data, and methods used. We then show the results and the econometric strategy to establish outcomes, and finally, we discuss the conclusions, limitations, and future work.

\section*{Methods}

\subsection*{Sample}

We collected data from 14 different schools and a total of 45 classrooms, between the 3rd and 5th grade. We analyzed data on 946 students with an average age of $10.16 \pm 1.13$. A descriptive table for each classroom can be found on Supplementary Methods 1. We also collected administrative records for each student, including classroom attendance, gender, educational level of student's parent or guardian, and GPA. Most of the students in the sample have been members of the same classroom for more than three years, spending together around eight hours each school day.

Our data collection methods and experimental protocol was approved on May 4, 2016 by the human subjects review board of Universidad del Desarrollo (IT15I10079). 

\subsection*{The Game}

\begin{figure}[t!]
\centering
\includegraphics[width=0.99\linewidth]{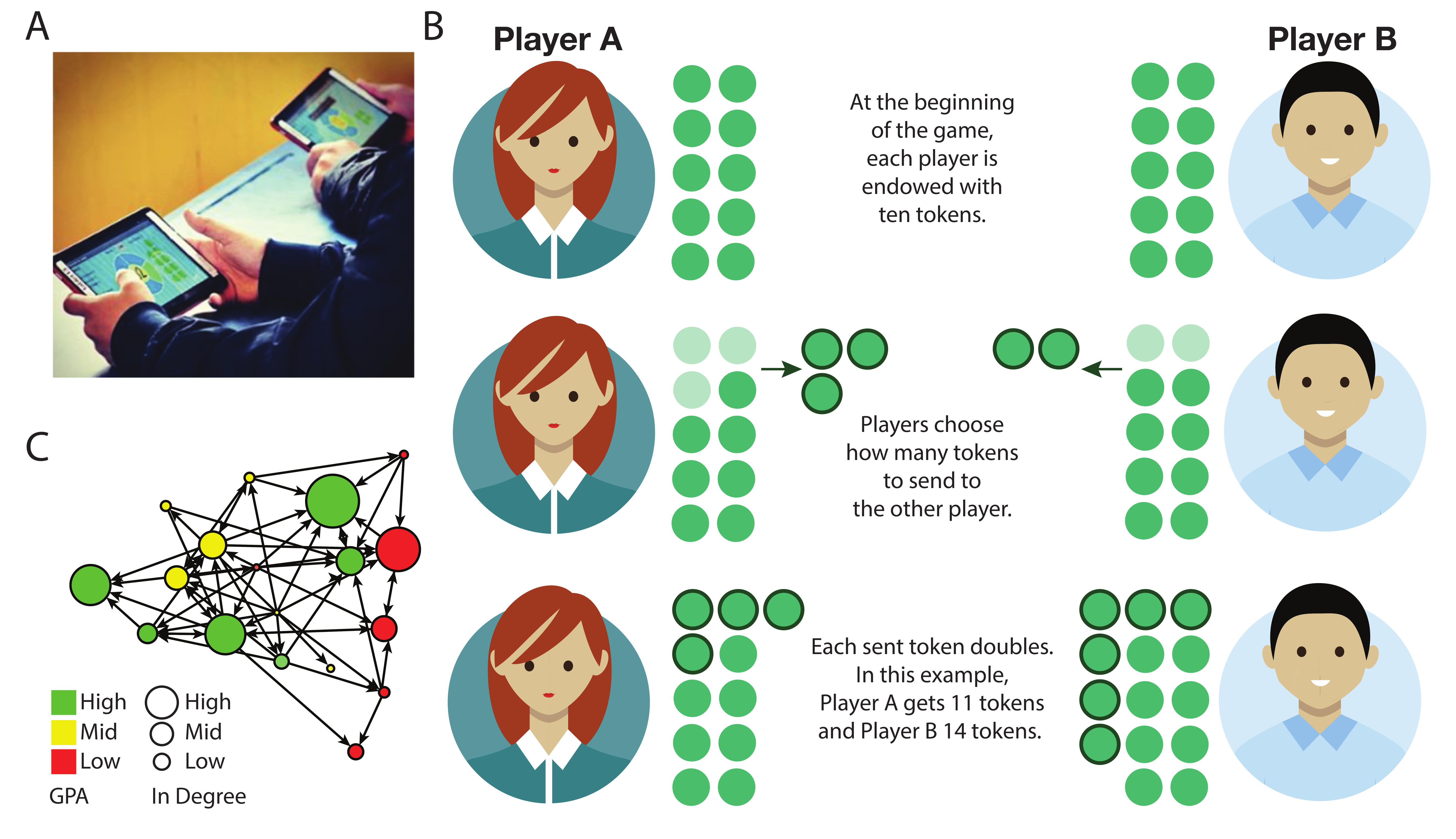}
\caption{Experimental game. A) Video game interface. B) Social dilemma: i) Both players are endowed with 10 (ten) tokens. ii) Simultaneously, player A decides to send 3 (three) tokens, and player B decides to send 2 (two) tokens. iii) Sent tokens are doubled. iv) Player A receives 4 (four) tokens and player B receives 6 (six) tokens. C) Students' network, edges represent fully cooperative interactions.}
\label{video_game}
\end{figure}

To measure cooperation, we implement a modified Prisoner's Dilemma using a friendly interface in tablet computers (Figure 1A). In this modified Prisoner's Dilemma, students play a non-anonymous simultaneous game once with each of their classmates (Figure 1B). In each dyadic interaction both students are endowed with ten tokens. Each student decides how many tokens to give to the other students. Every received token doubles, so students get the tokens they keep plus twice the tokens they receive. We use these cooperative interactions to map students networks. For instance, Figure 1C shows a network in which nodes represent students and directed links indicate fully cooperative interactions (a student giving ten tokens to another students).

\subsection*{Peer Nominations}

We also collected survey data using a nomination-based psychometric test \cite{Ryan2001}. This instrument collects information on socio-behavioral characteristics using peer-nominations focused on four dimensions: aggressive dispositions, prosociality, popularity, and social preferences (A description of the survey is available in Supplementary Methods 2).

\subsection*{Network measures}
We quantify individual's social capital using network measures. We define an adjacency matrix for each classroom as $w$, where each entry ($w_{ij}$) represents the number of sent tokens from student $i$ to student $j$. $N$ is the number of students in each classroom. Table \ref{measures} shows the equivalency between both, social capital and network metrics, besides it shows the mathematical expressions for each network measure.

\begin{table}[!h]\centering 
\begin{tabular}{lll}
Network measure     & Social Capital       & Formula\\
\hline
Average in-degree   & Average received cooperation & $r_i=\frac{1}{N}\sum_{j \neq i} w_{ji} $ \\
Average out-degree  & Average sent cooperation     & $s_i=\frac{1}{N}\sum_{j \neq i} w_{ij} $  \\
Reciprocated weight & Reciprocated cooperation     & $R_i=\frac{1}{N}\sum_{j\neq i}min[w_{ji},w_{ij}] $\\
PageRank           & Social ranking                  & $Rank_i =  \frac{1-d}{N} + d  \sum_{j=1}^n \frac{w_{ij}Rank_j}{\sum_{k=1}^n w_{kj}}$
\end{tabular}
\caption{Equivalency between network and social capital measures and their mathematical formulas. d represents a dumping factor, inventors of PageRank \cite{page1999pagerank} recommend to set it as $d=0.85$.}
\label{measures}
\end{table}

Evidence on the the external validity of game-theoretic experiments \cite{Karlan2005, Fehr2011, Gelcich2013, Hopfensitz2017, Algan2013} shows that real-life cooperation is correlated with cooperation under laboratory condition. Thus, we expect that these cooperative dispositions at the individual level will be revealed by the behavior implemented in the games. The main difference between standard experiments and our protocol comes from the non-anonymous feature of laboratory interactions. Thereby, we expect to capture not only the individual's intrinsic disposition to cooperate but their dispositions to cooperate within the context of a particular relationship and their history together \cite{Wang2017}.

\section*{Results}

\subsection*{Strategic behavior among students}

What strategies students implemented in these cooperative games? And how do these strategies differ between high GPA and low GPA students?

\begin{figure}[t!]
\centering
\includegraphics[width=0.9\linewidth]{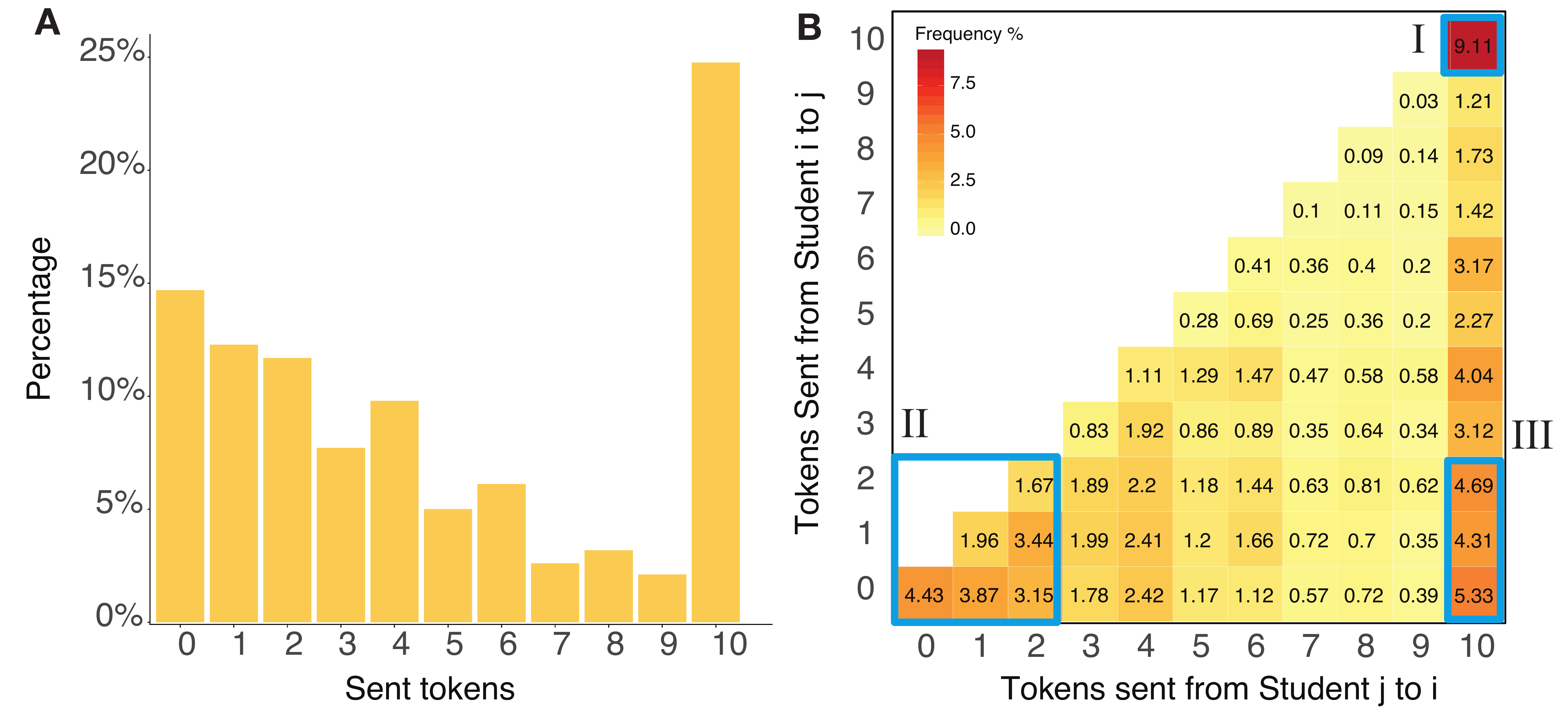}
\caption{Patterns of sent and received tokens across all dyadic interactions. A) Uni-variate distribution for sent tokens.  B) Bi-variate distribution of sent tokens. We observe three peaks describing the most commonly used strategies: I) Social optimum (top-right). II) Nash equilibrium (bottom-left). III)  Free-rider (bottom-right).}
\label{hists}
\end{figure}

Figure \ref{hists} A depicts an histogram summarizing the tokens sent and received by students across $18,334$ dyads. The figure shows that, despite the pseudo-continuity of the experimental design, students engage in strategies that are either fully cooperative or rather defective, sending mostly 10 or less than 2 tokens. Figure \ref{hists} B shows a bi-variate version of the histogram, showing that dyadic interactions concentrate close to the corners. In fact, more than $9\%$ of the interactions are fully cooperative (Figure \ref{hists} B I), while $18.7\%$ of interactions are highly defective, and involve both students sending two or less tokens to each other (Figure \ref{hists} B II). Asymmetric interactions are also visible in the game, with about $14.3\%$ of interactions involving a student sending ten tokens and getting two or less tokens in return (Figure \ref{hists} B III). Also, we find that students do send tokens across the entire range of values (Supplementary Figure 1), suggesting that the previous history of interactions among a pair of students matters for their decision-making process \cite{Wang2017}.

But, do students use different strategies depending on who they are playing with? While we observe that both low and high GPA students send, on average, a similar number of tokens ($4.7$ and $4.6$, respectively), high GPA students receive relatively more tokens ($5.0$ and $4.3$, respectively), indicating that high GPA students are more often the target of cooperation. But is this surplus in received tokes the result of a global preference to cooperate with high GPA students? Or is it the result of high GPA students being more strategic in the way they allocate their cooperative "investments"? 

\begin{figure}[t!]
\centering
\includegraphics[width=0.99\linewidth]{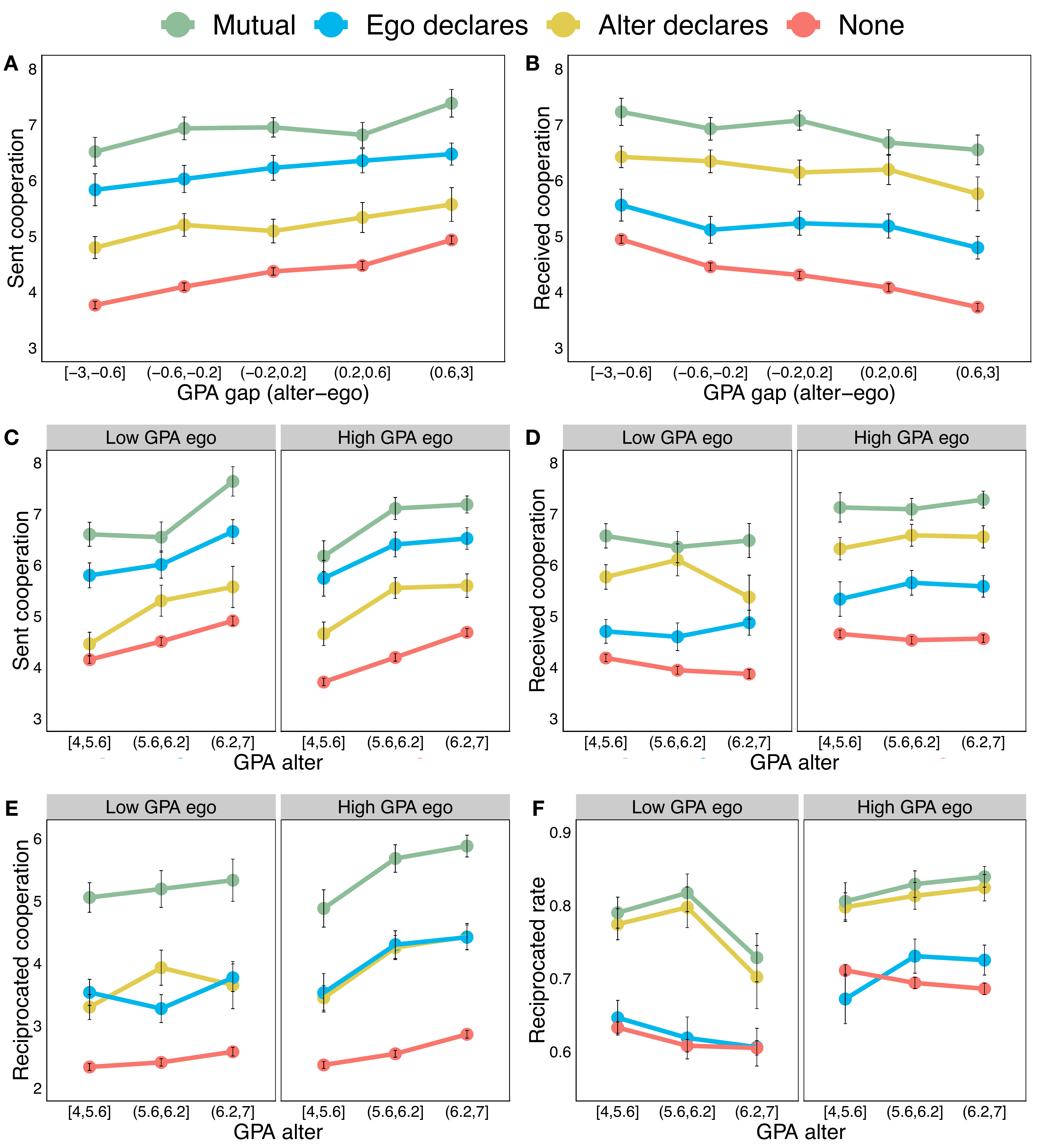}
\caption{Cooperative strategies of kids in elementary schools for high and low GPA student (GPA ego), considering declared friendship (colored lines).  A) Average sent cooperation. B) Average received cooperation. In panel A and B, x-axis represents the quintiles of the difference (gap) between alter and ego students GPA in each interaction. C) Average sent cooperation separated by ego's GPA. D) Average received cooperation separated by ego's GPA. E) Average reciprocated cooperation separated by ego's GPA. F) Average reciprocated rate (reciprocated cooperation over sent cooperation) separated by ego's GPA. In panels C, D, E, and F x-axis represents the terciles of alter students GPA in each interaction.}
\label{desc_1}
\end{figure}

Figure \ref{desc_1} shows different average metrics for cooperative behavior regarding student's GPA for four different friendship categories. Our results describe ego students behavior and alter students refer to students whom egos are interacting with. For instance, if student $i$ sent cooperation to student $j$, student $i$ is the ego student and $j$ is the alter student; if student $k$ receive cooperation from student $l$, student $k$ is the ego and $l$ is the alter. The friendship categories are: i) Both students, ego and alter, declare mutual friendship (green); ii) Only ego student declares friendship (blue); iii) Only alter student declares friendship (yellow); iv) No friendship declared (red).

Figure \ref{desc_1} A shows the average sent cooperation regarding the difference of GPA between alter and ego students, that is, the GPA gap. The GPA gap quantifies the difference between ego and alter's GPA. If GPA gap is positive, it means that alter's GPA is higher than ego's GPA. For instance, if the lowest GPA ego student sent cooperation to the highest GPA alter student, the GPA gap is the maximum. In these kind of cases, when the difference is positive and big, we say that the alter student is out of the ego's league. 

In figure \ref{desc_1} A we observe that students tend to send more cooperation to high GPA students, and this effect increases with the GPA gap, suggesting that there are cooperative investments in students out of the ego's "league." Figure \ref{desc_1} B shows the average received cooperation as a function of the GPA gap. We also observe that students receive lees cooperation from students out of the ego's league. 
 
But, do students with higher GPAs use different strategies than lower GPA students? Figure \ref{desc_1} C shows the cooperation sent by low and high GPA ego students (students below and above the median) to alters from different friendship categories sorted by their GPA. In absolute terms, we find that students send more tokens to mutual friends, followed by ego-friendships, and alter-friendships. Both high and low GPA students send the least amount of tokens to their non-friends. Yet, the effect of friendship is stronger for high GPA students because of two reasons. First, high and low GPA ego students send more cooperation to higher GPA alters (all t-tests comparing the highest tercile and the lowest tercile are significant with a p-value $<0.05$). Second, compared to low GPA ego students, high GPA ego students send less cooperation to their non-friends (t-test comparing non-friends of low GPA and high GPA egos has a p-value = $1.012e-05$), and statistically the same cooperation to their friends (green, blue, and yellow), meaning that higher GPA students are equally cooperative than low GPA students but only in mutual and asymmetrical friendship relationships.

Next, we explore received cooperation. If the sender aims to send cooperation that will be reciprocated from the receiver, they have to predict, based on their previous history, the amount of cooperation they will get back from each of their classmates. Figure \ref{desc_1} D shows two interesting patterns. First, relative to low GPA ego students, high GPA ego students obtain more cooperation across all friendship categories. Second, in all friendship categories for both, low and high GPA ego students, the received cooperation seems to follow a flat trend regarding alter students' GPA. This result suggests that, after accounting by the GPA of the ego, received cooperation does not vary with the GPA of the alter.

Sent and received cooperation, taken together, suggest that high GPA students are more strategic with their friends, sending more cooperation if mutual or asymmetrical friendship exists. To explore this , we define reciprocated cooperation \cite{squartini2013reciprocity}, $R_{i}$, as the overlap between the number of sent ($s_i$) and received tokens ($r_i$), $R_{i}=min[s_i, r_i]$ (Fig. \ref{desc_1} E). We note that reciprocated cooperation is highest for mutual friendship and lowest for non-friendship. Also, relative to each friendship category, high GPA ego students reciprocate more cooperation (from middle and high alter GPA students) than low GPA ego students.

To quantify the efficiency of the investments, we define the reciprocated rate as reciprocated cooperation over sent cooperation, $RoR_i=R_i/s_i$. Here, we observe that both, mutual friendship (green) and alter declared friendship (yellow), have the same rate of reciprocation (Fig. \ref{desc_1} F). At the same time, we observe that non-friendship (red) and ego declared friendship (blues) have the same rate of reciprocation, which is lower than the other two categories. Finally, relative to each friendship relationship, we observe that the lowest reciprocated rate (\ref{desc_1} F) is for low GPA egos sending cooperation to high GPA alters.This shows that high GPA alters tend to reciprocate more selectively, reciprocating more cooperation from high GPA alters than from low GPA alters. 

Our results reveal that elementary students are quite good at recognizing the directionality of friendship, and cooperate accordingly. Yet, high GPA students seem better at exploiting these asymmetries. 

\begin{table}[!t] \centering 
  
\scriptsize 
\begin{tabular}{@{\extracolsep{5pt}}lcccc} 
\\[-1.8ex]\hline 
\hline \\[-1.8ex] 
 & \multicolumn{4}{c}{\textit{Dependent variable:}} \\ 
\cline{2-5} 
\\[-1.8ex] & \multicolumn{4}{c}{GPA (after measuring)} \\ 
\\[-1.8ex] & (1) & (2) & (3) & (4)\\ 
\hline \\[-1.8ex] 
 Reciprocated cooperation  &  &  & 0.050$^{***}$ & 0.088$^{***}$ \\ 
 (z-score) &  &  & (0.015) & (0.033) \\ 
  & & & & \\ 
 Sent cooperation  &  &  &  & $-$0.057$^{**}$ \\ 
 (z-score) &  &  &  & (0.026) \\ 
  & & & & \\ 
 Rank  &  &  &  & 0.065$^{**}$ \\ 
 (z-score) &  &  &  & (0.026) \\ 
  & & & & \\ 
 Grades  & 0.720$^{***}$ & 0.746$^{***}$ & 0.732$^{***}$ & 0.699$^{***}$ \\ 
 (before measuring) & (0.024) & (0.023) & (0.024) & (0.025) \\ 
  & & & & \\ 
 Class attendance & 0.001 & $-$0.0004 & $-$0.001 & $-$0.0001 \\ 
 ($\%$)  & (0.001) & (0.001) & (0.001) & (0.001) \\ 
  & & & & \\ 
 Tutor comp. sec. school  & 0.038 & 0.028 & 0.027 & 0.024 \\ 
 (yes) & (0.026) & (0.024) & (0.024) & (0.024) \\ 
  & & & & \\ 
 Gender  & $-$0.089$^{***}$ & $-$0.063$^{**}$ & $-$0.056$^{**}$ & $-$0.053$^{**}$ \\ 
 (Male) & (0.026) & (0.027) & (0.027) & (0.026) \\ 
  & & & & \\ 
 Class size & 0.0002 &  &  &  \\ 
  & (0.002) &  &  &  \\ 
  & & & & \\ 
 Constant & 1.448$^{***}$ &  &  &  \\ 
  & (0.162) &  &  &  \\ 
  & & & & \\ 
\hline \\[-1.8ex] 
Fixed effects & Pooling & Class & Class & Class \\ 
Observations & 769 & 769 & 769 & 769 \\ 
R$^{2}$ & 0.613 & 0.636 & 0.641 & 0.652 \\ 
Adjusted R$^{2}$ & 0.610 & 0.612 & 0.617 & 0.627 \\ 
F Statistic & 241.733$^{***}$ & 314.478$^{***}$  & 257.097$^{***}$  & 191.879$^{***}$  \\ 

 & (df = 5; 763) & (df = 4; 720) &  (df = 5; 719) & (df = 7; 717) \\ 
\hline 
\hline \\[-1.8ex] 
\textit{Note:}  & \multicolumn{4}{r}{$^{*}$p$<$0.1; $^{**}$p$<$0.05; $^{***}$p$<$0.01} \\ 
\end{tabular} 
  \caption{OLS regressions for students' GPA after measuring cooperation.} 
\label{table1} 
\end{table}

\subsection*{Identification Strategy}

Do students that participate in a high number of mutually cooperative relationships increase their GPAs over time faster than other students? 

To study the relationship between reciprocity and GPA improvement, we set up a statistical model (eq. \ref{eq_model1}) by estimating the individual future GPA as a function of the individual average reciprocated cooperation (Table \ref{table1}). Model 1 sets the baseline, where we include classic confounding variables such as gender ($G_i$), percentage of class attendance ($A_i$), level of education of the guardian ($TESC_i=1$ if the guardian completed secondary school, 0 otherwise), and the number of students in each classroom ($CS_i$). 

In models 2 to 4, we control by the classrooms average unobserved differences using fixed effects $\theta_i$, here we note how class size variable is absorbed by classrooms fixed effects. In model 3, we observe the effect of the individual average reciprocated cooperation $R_i$ (presented as z-score in the whole sample for interpretability). 

Finally, model 4 includes three elements: i) Individual average sent cooperation, $s_i$, it quantifies the student perception of others; ii) Individual-level social ranking, $Rank_i$, it is calculated as the page rank \cite{page1999pagerank}, and it quantifies the social perception of individuals as a ranking \cite{bruch2018aspirational}, considering the ego's received cooperation and the social ranking of the alter students that send cooperation to the ego student; and iii) Previous semester's GPA, $GPA_{i0}$, it is a proxy for individual talent and it also controls by several confounding variables that impact GPA, such as, household income, practicing sports, among others. These three elements are presented as z-score over the whole sample for interpretability. Our specification is the following:

\begin{equation}
GPA_{i1} = \beta_1 R_i + \beta_2 s_i + \beta_3 Rank_i + \beta_4 GPA_{i0}+\beta_5 G_{i} + \beta_6 A_i +\beta_7 TESC_i + \beta_8 CS_i + \theta_c+ e_i, \label{eq_model1}
\end{equation}
where, $GPA_{i1}$ represents the future GPA of the student $i$ and $e_i$ is the error term.

We find a positive and significant effect for reciprocated cooperation. More precisely, we find that an increase of one standard deviation in reciprocated cooperation is associated with an increase of $0.088$ units in future GPA. We note that GPA in Chile goes from $1.00$ to $7.00$, and the average GPA of the first and the second semester of 2017 are $5.87$ and $5.77$, respectively. Thus, the average variation between both semesters is $\Delta_{GPA}\approx -0.10$, therefore, the reciprocated cooperation effect size ($0.088$) and the average decreasing of GPA between the two periods ($0.100$) are comparable.

Nevertheless, some meaningful changes could be affecting our results. For instance, changes in classroom's configuration such as new teachers and new students in the classroom or increasing/decreasing in the school income. Also, changes at the individual level, such as, a new job for the student's guardian that would impact household income, house moving that would impact student's social capital outside of the classroom. Other individual changes, such as, a student's ill, a trip, failing school year, or family issues would impact school engagement, among others. 

However, we assume that most of these unobserved variables are time-invariant within the time span of our study. To evaluate the plausibility of this assumption, it is important to have in mind some socio-economical configurations of Chile. First, from elementary to high school, students remain together with their classmates in all classes the whole year (and even the whole school cycle), i. e., students spend together around 8 hours per week, five days a week from March to December each year of the school, cycle. Second, given the academic year starts in March, it is completely overlapped with the fiscal year of Chile (which starts on January 1st). Thus, both, the socio-economical configurations in Chile and our data collection design, which spans within the year 2017, allow us to plausibly assume that classroom and school-level changes are very rare. At the individual level, student's guardian job changes, and student's housing moving have very low probabilities of occurrence within the fiscal year, mostly because all of these changes happen to occur in holidays, between December and March, quite motivated by economical reasons. Finally, we can reasonably assume that some of the variables impacting student's engagement remain time-invariant, such as long trips or failing the school year (this decision is taken by the school at the end of the school year, December). However, illnesses and family issues are not constrained to systemic configurations, and they impact directly to class attendance, that is why we control for class attendance in both observed periods. Finally, our data comes from vulnerable public schools on a peripheral county of the capital province of Chile, then we can assume a certain level of homogeneity in household incomes. Moreover, we are controlling by guardian's level of education that is a proxy for household income, therefore, we are also capturing income variations. 

\begin{table}[!t] \centering 
\tiny 
\begin{tabular}{@{\extracolsep{5pt}}lcccccc} 
\\[-1.8ex]\hline 
\hline \\[-1.8ex] 
 & \multicolumn{6}{c}{\textit{Dependent variable:}} \\ 
\cline{2-7} 
\\[-1.8ex] & \multicolumn{6}{c}{GPA }  \\ 
\\[-1.8ex] & \multicolumn{6}{c}{(Diff-in-Diff)} \\
\\[-1.8ex] & (1) & (2) & (3) & (4) & (5) & (6)\\ 
\hline \\[-1.8ex] 
 Reciprocity * Time & 0.047$^{*}$ & 0.043$^{*}$ & 0.043$^{*}$ & 0.044$^{*}$ & 0.039$^{***}$ &  \\ 
  & (0.027) & (0.026) & (0.025) & (0.023) & (0.012) &  \\ 
  & & & & & & \\ 
 Dummy reciprocity * Time &  &  &  &  &  & 0.084$^{***}$ \\ 
  &  &  &  &  &  & (0.027) \\ 
  & & & & & & \\ 
 Reciprocated cooperation & 0.096$^{***}$ & 0.061$^{***}$ & 0.139$^{***}$ & 0.265$^{***}$ &  &  \\ 
  (z-score)& (0.019) & (0.018) & (0.025) & (0.034) &  &  \\ 
  & & & & & & \\ 
 Time & $-$0.083$^{***}$ & $-$0.108$^{***}$ & $-$0.107$^{***}$ & $-$0.106$^{***}$ & $-$0.092$^{***}$ & $-$0.113$^{***}$ \\ 
  & (0.026) & (0.025) & (0.024) & (0.022) & (0.012) & (0.013) \\ 
  & & & & & & \\ 
 Sent cooperation  &  &  & $-$0.128$^{***}$ & $-$0.206$^{***}$ &  &  \\ 
 (z-score) &  &  & (0.020) & (0.025) &  &  \\ 
  & & & & & & \\ 
 Rank  &  &  & 0.107$^{***}$ & 0.106$^{***}$ &  &  \\ 
 (z-score) &  &  & (0.020) & (0.027) &  &  \\ 
  & & & & & & \\ 
 Class attendance &  & 0.020$^{***}$ & 0.019$^{***}$ & 0.018$^{***}$ & 0.006$^{***}$ & 0.006$^{***}$ \\ 
 ($\%$)  &  & (0.001) & (0.001) & (0.001) & (0.001) & (0.001) \\ 
  & & & & & & \\ 
 Tutor comp. sec. school&  & 0.145$^{***}$ & 0.130$^{***}$ & 0.121$^{***}$ &  &  \\ 
   (yes) &  & (0.025) & (0.024) & (0.024) &  &  \\ 
  & & & & & & \\ 
 Gender  &  & $-$0.213$^{***}$ & $-$0.224$^{***}$ & $-$0.129$^{***}$ &  &  \\ 
 (Male)&  & (0.025) & (0.025) & (0.026) &  &  \\ 
  & & & & & & \\ 
 Class size &  & 0.001 & 0.011$^{***}$ &  &  &  \\ 
  &  & (0.002) & (0.003) &  &  &  \\ 
  & & & & & & \\ 
 Constant & 5.908$^{***}$ & 4.074$^{***}$ & 3.983$^{***}$ &  &  &  \\ 
  & (0.018) & (0.131) & (0.131) &  &  &  \\ 
  & & & & & & \\ 
\hline \\[-1.8ex] 
Fixed effects & Pooling & Pooling & Pooling & Class & Individual & Individual \\ 
R$^{2}$ & 0.048 & 0.216 & 0.259 & 0.338 & 0.905 & 0.904 \\ 
Adjusted R$^{2}$ & 0.047 & 0.213 & 0.255 & 0.326 & 0.828 & 0.827 \\ 
Observations & 1,866 & 1,676 & 1,676 & 1,676 & 1,676 & 1,676 \\ 
\hline 
\hline \\[-1.8ex] 
\textit{Note:}  & \multicolumn{6}{r}{$^{*}$p$<$0.1; $^{**}$p$<$0.05; $^{***}$p$<$0.01} \\ 
\end{tabular} 
   \caption{Difference-in-difference estimation of the reciprocated cooperation effect, and first differences estimation. 1) In model 5, we build a dummy variable for reciprocated cooperation, it takes value 1 if the reciprocated cooperation for individual $i$ is in the top quartile, and it takes value 0 otherwise. 2) In model 7, the coefficient for reciprocated cooperation should be statistically the same as the diff-in-diff estimator (reciprocated cooperation and time interaction) for model 1 to 5. 3) In model 7, the constant should be statistically the same as the time effect for models 1 to 5. 4) The time invariant controls and fixed effects are canceled by first differences.} 
  \label{table11}   
\end{table}

To overcome all of these issues and to provide evidence on the magnitude of the relationship between reciprocated cooperation and GPA improvement, we use a treatment intensity difference-in-difference framework \cite{acemoglu2004women}. Here, our treatment intensity variable is individual average reciprocated cooperation ($R_i$), a continuous variable, which induces variation at the individual level. Our specification is the following:

\begin{equation}
GPA_{it}=\beta_{1}+\beta_{2}T+\beta_{3}R_i+\beta_4 A_{it} +\beta_5 X_i+\delta T \times R_i+\varepsilon_i+ e_{it},
\label{eq_model2}
\end{equation}
where $\varepsilon_i$ represents individual-level fixed effects (we note that individual-level fixed effects absorb classroom level fixed effects) and  $e_{it}$ is the error term. $GPA_{it}$ is the GPA of student $i$ in period $t$, $T$ represents the semester and it takes values 0 (before measuring, $t=0$) and 1 (after measuring, $t=1$), $R_i$ is the treatment intensity (reciprocated cooperation), $A_{it}$ is the class attendance for both time periods, $X_i$ is a vector of time-invariant individual controls, and finally, the diff-in-diff estimator is represented by $\delta$. Table \ref{table11} shows six variations of the specification showed in eq. \ref{eq_model2} and model 7 shows the equivalent first difference strategy. In models 1-5 we observe that our diff-in-diff estimator is robust under different sub-specifications of eq. \ref{eq_model2}, even after controlling by individual fixed effects (model 5). We note that fixed effects in model 5 absorb all the individual and classroom level controls. Both remaining coefficients, time and the interaction between reciprocated cooperation and time, are quite consistent across all the models. In model 6 we use a dummy variable for the treatment by setting the dummy as $1$ for all the individuals in the top quartile of reciprocated cooperation and $0$ otherwise. Given the nature of dummy variables, we observe a small increase in the parameters, however, the errors also increase, in fact, there is error overlapping between model 6 and all the previous models.

Therefore, we argue that our diff-in-diff estimator is consistent and robust across all the models (Table \ref{table11}  models 1-6).

Now, let us focus on model 5, the diff-in-diff estimator is positive ($0.039$) and significant (p-value$<0.01$), even after controlling by all the unobserved time-invariant individual characteristics, which is approximately a half of the associative effect presented in Table \ref{table1} model 4. The average GPA difference between $t=0$ and $t=1$ for controls and counterfactual are $0.100$ GPA units. 

Here, we intended to provide evidence on the effect of reciprocated cooperation on GPA improving. The central assumption is that both control and treatment groups follow parallel trends in the absence of our treatment intensity variable, the reciprocated cooperation. Changes in GPA follow the same trend in the absence of reciprocated cooperation for all students. Some limitations exist here; for instance, if, within the time interval studied, any change in a confounding variable that affects both, GPA and reciprocated cooperation, could hinder our results. For example, if between the first and second semester of 2017 a change in the level of a guardian's student income or a student started to practice sports, could have impacted both, reciprocated cooperation and GPA. These changes would make the parallel trend assumption unfulfilled because the initial GPA would control none of these changes. Income may impact reciprocity through the popularity of a student, and also it may affect GPA through getting a private professor or getting access to internet \cite{Jackson2006}. Also, practicing sports may impact reciprocated cooperation through popularity, and also it may impact GPA \cite{Singh2012,rees2010sports}. 

Therefore, regardless of the consistency of our estimators and the reasonable assumptions in our specifications, we acknowledge that causal identification is always a quite difficult issue to address. Indeed, it is impossible to ensure that parallel trends assumption is wholly fulfilled. In this work, we intend to provide a piece of causal evidence on how reciprocated cooperation impacts GPA. However, to establish a more comprehensive casual relationship, more data have to be collected, and more work in this context has to be done.

\section*{Discussion}

The literature on learning states a positive relationship between a student's position in their social network and their academic performance. In this context, there are two possible explanations: i) Receiving cooperation leads to better GPAs through positive learning externalities; or ii) Higher GPA leads to higher status, resulting in more cooperation from students with lower GPA. Here, we find evidence against the latter by controlling by the previous student GPA, as a proxy of students' talent. Since the effect survives even after controlling by other confounding variables, such as gender, class attendance, and the educational level of the student's guardian, we find evidence in support of reciprocated cooperation as future predictor of GPA. 

A limited ability to recognize friendship ties, and their directionality, limits people's ability to engage in cooperative arrangements \cite{Almaatouq2016}. Indeed, people are typically bad at perceiving the directionality of friendship \cite{Almaatouq2016}. For instance, in a friendship nomination study on adolescents, only $64\%$ of the reported friendships were reciprocal \cite{vaquera2008you}. Our experimental set-up reveals that elementary school students are quite good at recognizing friendship because they cooperate accordingly to each type of friendship. 

To sum up, our results show that high GPA students benefit more than low GPA students from the underlying cooperative network in which learning is embedded. The advantage of high GPA students comes from two different channels. First, due to their academic status, high GPA students receive more cooperation than low GPA students. Second, high GPA students cooperate more selectively with their friends. To the extent that investing in friends increases the reciprocity of the relationships. This selectivity translates into a more efficient strategy to reap the learning benefits from their social networks. We provide evidence of a directional effect from reciprocated cooperation to GPA improving, under the assumption of parallel trend is fulfilled. We understand that parallel trend assumption is strong, we also understand that the underlying complexities behind social behavior are difficult to isolate in an experiment of this nature. However, we consider this work to contribute to our understanding of the link between social networks and learning outcomes. 

From a policy perspective, the natural question to ask is what kind of intervention might improve the academic performance of students by optimizing the potential benefits of cooperation. For instance, encouraging friendships through interventions of the spatial arrangement of the class \cite{Salend1999} or an instructional design that promotes the formation of social ties in the classroom \cite{11_McFarland,Rohrbeck2003, slavin2015cooperative, newman2002self, slavin2003cooperative,slavin2011instruction, pulgar2019physics} are good alternatives to explore. 

From a methodological perspective, our results open new avenues for the role of network-based tools to leverage social capital information in elementary education \cite{Burt2000, Lakon2008}.

\section*{Software Availability}
A scalable version of the video game, used to gather the data, can be found at \url{http://www.juecoo.udd.cl}.
Any school can log in and use the game to get an automatic report about their students. At this point, only a Spanish version of the video game is available, an English version will be launched soon.

\section*{Data Availability}
The anonymized data can be delivered under a reasonable request to the authors.

\section*{Acknowledgments}
The authors acknowledges the financial support from FONDEF IT15I10079. The authors are also thankful to Tamara Yaikin and Cecilia Monge for useful insights on the video game conception and design. The authors thanks to Nursoft for supporting in the video game implementation. Finally, the authors are also thankful to Javier Pulgar, Flavio Pinheiro, Cristian Jara-Figueroa, Gustavo Castro-Dominguez, Centro de Investigación en Complejidad Social (CICS), and Collective Learning Group at the MIT Media Lab, for the helpful insights and discussions.

\section*{Author Contributions}
C.C. contributed to data analysis.
C.C. and C.R-S. contributed to the study conception and design, interpretation of data, and writing the manuscript.
V.L-T. contributed to the study design, acquisition and data analysis.
C.A.H. contributed to the study conception and writing the manuscript.

\section*{Corresponding Authors}
Cristian Candia (cristian.candia@kellogg.northwestern.edu | ccandiav@mit.edu | ccandiav@udd.cl). \\
Carlos Rodriguez-Sickert (carlosrodriguez@udd.cl).

\bibliographystyle{ieeetr}
\bibliography{bib}

\pagebreak

\section*{Supplementary Video}
 An explanatory video about the platform and the game can be found at \url{https://youtu.be/jiFa58-Lugk}. English subtitles are available.

\section*{Supplementary Notes}

\subsection*{SM 1.1: What is Video-Game Theory?}
We conducted a lab in the field experiment. We implement a dyadic social dilemma, specifically a modified version of the Dictator game (Forsythe, 1994) in which both players simultaneously perform the role of the “Allocator” and the role of the “Recipient”. We also add a multiplicative factor on donations to allow for potential efficiency gains. Under standard game-theoretic experimental protocols, which involve anonymous interaction, networks elicited in the lab emerge from scratch, mainly through assortative interaction between anonymous players (see, for instance, Goeree et al, 2009). However, in order to capture the nature of pre-existing relationships we have to depart from the standard protocol and consider onymous interactions (Wang et al, 2017), i.e. in our social dilemma, students are aware of the identity of the partner that has been matched to them each round. Thereby, their choices when playing again each other are not only the result of intrinsic prosocial dispositions (or their absence), but also the result of their history and the perceptions they have about each other. 

To the extent that learning is a collective enterprise (Henrich, 2015) and several studies show the
importance of peer effects on school performance (for a survey on these results, see Sacerdote, 2011), we
expect that the position of the individual in the network together with the strength of her/his ties, could
modulate the capacity of the student to capture the externalities generated by their peers during the
learning process and, thus, influence her/his academic performance as suggested by previous theoretical
work (Calvo-Armengol and Patacchini, 2009).

Particularly, in elementary school sample, in each round every student play with a different classmate, in order to map a complete network of relations inside the classroom. Students are endowed with 10 tokens per round, that they can share or keep (cooperate or defect). The total number of tokens that a student gets in a round is equal to the number of tokens they kept plus twice the number of tokens they received. Hence, the dyad maximizes the total number of tokens earned when everyone cooperates, but students maximize their tokens when they defect and everyone else cooperates.

\subsection*{SM 1.2: Methodological challenges and alternative approaches}

In order to elicit the mutual support social network of a given class, we implement a dyadic social dilemma among all of the possible dyads of such class. The experiment recreates social exchange and forces an action with (virtual) consequences towards her/his counterparts. This feature of our measurement instrument might mitigate social desirability biases prevalent in declarative responses of survey-based instruments. We as economists have our own bias towards revealed preferences as a more reliable source of information than declared preferences (Becker, 2013). However, it is important to acknowledge that peer-nomination surveys allow for the explicit declaration of different categories of relationships, e.g., identify friends and also obtain information about the set of attributes that a student has as declared by her/his school mates, and thus, identify the most popular/aggressive students according to their peers.

To the extent that we also collect peer-nomination data, we can explore questions involving both sources of relational data. If the outcome of the dyadic game can predict declarations of mutual friendships in the peer nomination instrument, we can understand how gradual measures of the tie’s strength and the level of symmetry of such friendship correlate with such declarations. Goeree et al (2010), for instance studied the relationship between donations towards a given “recipient” and her/his position on the friendship network. The information collected about the set of attributes relevant in the context of social life (popularity, aggressiveness) might also help us to improve our understanding of the relationship between experiment-based elicitation of the topological properties of the student’s position in the network and her/his academic performance. High status individuals identified in the elicited network are built on the basis of prestige (popularity) or dominance (aggressiveness). The literature on social learning proposes that people are more likely to learn from those who are seen as prestigious, talented, or that share demographic attributes with learners, not just in adults but also within children (Chudek et al, 2012). Moreover, they propose a mechanism to explain how prestige can promote cooperation and create prosocial leaders (Henrich et al, 2015). 

In order to boost engagement and mitigate cognitive barriers (Camerer, 2003) associated with the young age of our sample and the vulnerable character of a big portion of our sample, we forego the use of standard experimental interfaces such as z-tree and develop a videogame-like didactical and friendly interface implemented using tablets.

\subsection*{SM 1.3: Benefits of Video-Game Theory}

\subsubsection*{Contribution to the external validity of game-theoretic experiments in school context}

We extend the experimental economics literature that use classrooms population involving social dilemmas (Fehr et al, 2008; Cárdenas et al, 2012), studying the relationship between experimental results and school outcomes, such as academic performance.

\subsubsection*{Relational focus}
Rather than focusing on the group or individual levels of cooperation (the social capital of a group and the individual’s contribution to the social capital of the group, respectively), we produce a measure of the individual’s social capital embodied directly in the agent’s relationships and indirectly by the configuration of the network as a whole (Lakon et al, 2008).

\subsubsection*{Novel link between experimental game theory and social networks}
Under standard game-theoretic experimental protocols, which involve anonymous interaction, networks elicited in the lab emerge from scratch, mainly through assortative interaction between anonymous players (see, for instance, Goeree et al, 2009). In this study, we are able to elicit a representative social network for a group of people with a common history, using methods from experimental game theory.

\subsubsection*{Novel approach to elicit the social network of mutual support in school communities}
The use of a social dilemma experiment implemented in the classroom and used as a measurement instrument, differentiates ourselves from standard approaches to elicit the mutual support network, which are built on the basis of peer nomination using survey-methods (Blansky et al, 2013). Moreover, to the extent that we also implement a standard peer-nomination instrument, we will be able to contrast both approaches and explore possible complementarity.

\pagebreak

\section*{Supplementary Methods}

 \subsection*{SM 2.1: Classrooms descriptions}
\begin{table}[!htbp] \centering 
  \caption{Class-level description.} 
\label{calss_desc}
\tiny
\begin{tabular}{lcccccccc}
\hline\hline

Class & Ave. Cooperation & Ave. Bullying & Ave. Victimization & Ave. Figthing & dm\_in & rec & p\_coop\_c & p\_dm\_in\\
\hline
\hline
School 1 3º & 52.9 & 35.4 & 47.5 & 34.8 & 0.1 & 0 & 4.3891304 & -0.05\\
School 1 4º & 47.6 & 21.1 & 46.7 & 18.9 & 0.1 & 0 & -0.9108696 & -0.05\\
School 1 5º & 44.0 & 23.4 & 46.0 & 23.7 & 0.1 & 0 & -4.5108696 & -0.05\\
\hline
School 2 3ºA & 60.5 & 29.2 & 40.7 & 31.0 & 0.1 & 0 & 11.9891304 & -0.05\\
School 2 3ºB & 48.1 & 13.3 & 36.0 & 17.5 & 0.1 & 0 & -0.4108696 & -0.05\\
School 2 4ºA & 50.7 & 27.6 & 47.0 & 15.7 & 0.2 & 0 & 2.1891304 & 0.05\\
School 2 5ºA & 42.5 & 27.1 & 46.6 & 30.6 & 0.2 & 0 & -6.0108696 & 0.05\\
School 2 5ºB & 72.6 & 27.0 & 51.4 & 21.9 & 0.2 & 0 & 24.0891304 & 0.05\\
\hline
School 3 3º & 59.1 & 27.4 & 35.0 & 26.7 & 0.1 & 0 & 10.5891304 & -0.05\\
School 3 4º & 48.8 & 35.2 & 40.5 & 20.5 & 0.2 & 0 & 0.2891304 & 0.05\\
School 3 5º & 51.6 & 19.2 & 35.3 & 16.1 & 0.2 & 0 & 3.0891304 & 0.05\\
\hline
School 4 4ºA & 53.1 & 25.6 & 57.2 & 16.5 & 0.1 & 0 & 4.5891304 & -0.05\\
School 4 4ºB & 43.1 & 24.2 & 58.8 & 16.7 & 0.1 & 0 & -5.4108696 & -0.05\\
School 4 5ºA & 34.1 & 25.5 & 48.4 & 26.7 & 0.1 & 0 & -14.4108696 & -0.05\\
\hline
School 5 3º & 62.3 & 24.8 & 50.8 & 18.7 & 0.2 & 0 & 13.7891304 & 0.05\\
School 5 4º & 58.5 & 24.5 & 52.8 & 18.3 & 0.3 & 0 & 9.9891304 & 0.15\\
School 5 5º & 40.9 & 34.8 & 50.0 & 32.0 & 0.2 & 0 & -7.6108696 & 0.05\\
\hline
School 6 3º & 45.0 & 27.1 & 51.8 & 26.0 & 0.1 & 0 & -3.5108696 & -0.05\\
School 6 4º & 60.5 & 16.3 & 32.1 & 17.5 & 0.1 & 0 & 11.9891304 & -0.05\\
School 6 5º & 46.6 & 20.1 & 39.9 & 21.1 & 0.1 & 0 & -1.9108696 & -0.05\\
\hline
School 7 3º & 49.4 & 18.4 & 31.7 & 20.3 & 0.2 & 0 & 0.8891304 & 0.05\\
School 7 4º & 40.0 & 27.6 & 66.2 & 24.3 & 0.2 & 0 & -8.5108696 & 0.05\\
School 7 5º & 43.6 & 25.6 & 46.2 & 28.8 & 0.2 & 0 & -4.9108696 & 0.05\\
\hline
School 8 3º & 65.8 & 24.1 & 52.2 & 23.2 & 0.2 & 0 & 17.2891304 & 0.05\\
School 8 4º & 55.2 & 12.8 & 34.4 & 14.2 & 0.2 & 0 & 6.6891304 & 0.05\\
School 8 5º & 51.9 & 28.1 & 39.2 & 22.7 & 0.2 & 0 & 3.3891304 & 0.05\\
\hline
School 9 3ºA & 42.6 & 23.0 & 54.0 & 26.7 & 0.2 & 0 & -5.9108696 & 0.05\\
School 9 3ºB & 62.8 & 20.6 & 25.0 & 22.7 & 0.1 & 0 & 14.2891304 & -0.05\\
School 9 4ºA & 37.1 & 26.2 & 47.9 & 29.7 & 0.1 & 0 & -11.4108696 & -0.05\\
School 9 4ºB & 43.0 & 21.1 & 39.2 & 17.2 & 0.1 & 0 & -5.5108696 & -0.05\\
School 9 5ºA & 35.1 & 35.8 & 60.4 & 37.8 & 0.1 & 0 & -13.4108696 & -0.05\\
\hline
School 10 3º & 62.5 & 26.4 & 49.0 & 32.9 & 0.2 & 0 & 13.9891304 & 0.05\\
School 10 4º & 38.4 & 24.7 & 51.4 & 36.1 & 0.2 & 0 & -10.1108696 & 0.05\\
School 10 5º & 34.6 & 26.5 & 65.3 & 28.9 & 0.2 & 0 & -13.9108696 & 0.05\\
\hline
School 11 3º & 50.2 & 20.0 & 56.1 & 24.0 & 0.1 & 0 & 1.6891304 & -0.05\\
School 11 4º& 48.1 & 23.8 & 46.5 & 26.1 & 0.2 & 0 & -0.4108696 & 0.05\\
School 11 5º& 32.9 & 26.4 & 52.5 & 27.8 & 0.1 & 0 & -15.6108696 & -0.05\\
\hline
School 12 4A & 39.0 & 7.7 & 39.1 & 8.5 & 0.1 & 0 & -9.5108696 & -0.05\\
School 12 4B & 50.6 & 15.8 & 42.6 & 16.3 & 0.1 & 0 & 2.0891304 & -0.05\\
School 12 4C & 54.5 & 18.0 & 32.6 & 9.9 & 0.1 & 0 & 5.9891304 & -0.05\\
\hline
School 13 3ºA & 51.0 & 13.2 & 34.1 & 9.5 & 0.1 & 0 & 2.4891304 & -0.05\\
School 13 3ºB & 59.0 & 16.7 & 39.0 & 13.6 & 0.2 & 0 & 10.4891304 & 0.05\\
School 13 4º & 49.9 & 9.6 & 30.7 & 12.7 & 0.2 & 0 & 1.3891304 & 0.05\\
School 13 5º & 41.5 & 14.2 & 40.8 & 15.1 & 0.2 & 0 & -7.0108696 & 0.05\\
\hline
School 14 4º & 40.7 & 29.3 & 54.1 & 27.6 & 0.1 & 0 & -7.8108696 & -0.05\\
\hline
\hline
\end{tabular}
\end{table}

\pagebreak

\subsection*{SM 2.2: Fieldwork protocol, FAQ and game instructions}

\subsubsection*{Protocol of fieldwork implementation}

Fieldwork teams must implement this protocol in each occasion. If specific circumstances force fieldwork team to make any modification, these changes must be informed in Field Report notes (see annex) as protocol observation.

Stage 0: Previous actions to school implementation of the game: 
To verify that activity schedule (day and time) can be carried out.
To verify that number of tablets and headphones is enough for the activity.
To verify that each class data is right entered in tablets.
To review the list of students and to register those absent students in fieldwork day.
To turn on tablets and to verify their connection to the system and that their battery charge is enough for implementation.
To prepare the set of materials to carry out the activity:
Data show and speakers
Computer for teacher survey 
Computer for the videogame management
Tablets (At least the number of students plus 3 for replacement)
Headphones (At least the number of students plus 3 for replacement)
Router in case of no internet in school

Stage 1: Classroom preparation 
To verify together with the teacher that any of the students in the room has not been allowed by their parents to participate. If the number of absent and not allowed to participate students is 25 per cent or more of the class, the game implementation must be cancelled. 
To pile up the tablets on teacher table or other appropriate desk. Tablets must not be distributed to students until game presentation is over. Do not distribute tablets until the game starts on.

To put the desks apart in classroom so each student is seated alone (see figure below). If the school desks are for two or more students and it is not possible to separate them fieldwork team must register it as an observation in the implementation protocol.

To enumerate desks beginning in the first row (nearest to board) from right to left and finalizing in the last row, as shown next:

If there are backpacks or other objects between the desks they must be put leaning against the wall before beginning the activity. 

Stage 2: Students income and setup

If activity is implemented just after recess students must be asked to make a row in list order and to take seat on the desks in correlative order. First student in the row must seat in desk 1, second student on desk 2 and so on. 
Students are invited to come into the classroom and to take seat on desks in correlative order according they go into the room.
Each student must be asked to clear the desk and to put away everything that is over it.
When all students are seated, fieldwork team must welcome them and read the instructions for the activity according to next statements:
“Hello, welcome everyone!
We are from Universidad del Desarrollo. We bring these tablets that you can see here. They contain a game that we would like to share with you. Do you want to play?
Well, we will distribute a set of one tablet and headphones to each one of you. When you receive it you must put the tablets face down on your desks until we tell you to turn it around and put headphones on. But, after we begin, I am going to talk you about the game”.

Stage 3: Distribution of tablets and instructions for surveys
Fieldwork team distributes tablets face down and headphones connected to them. Students are asked not to turn tablet around until everybody has his own.
Once every student has received a tablet, the fieldwork attendant says: “Now we are almost ready to start the game. There is only one thing left. Before we start we want you to respond some questions that will be in the screen of your tablet. There will be two surveys. It is important you to know that there are no right or wrong answers. Feel free to answer what you think. First survey’s questions are about you. We will wait until everybody is ready to move on to the second survey about your class, right? Remember, first you are going to answer on tablets a survey about yourselves, when you finish you must wait everybody to be ready and I will tell you when you can begin with the second survey about your class. After all students have finished both surveys I will talk you about the games, ok? (DO NOT READ: Authorize the start of instructions on the server). 
You can turn your tablets around and answer the questions. Remember to stay in silence”.
Instructors walk around the classroom assisting and answering doubts.
To verify that every students have their headphones on.

Stage 4: Game instructions
	
“Very well, now we are going to see what the game is about. We have prepared a short video that explains the game”.

Couples game instructional video is shown. Link to video (english subtitles not available yet).

“The Couples game is very simple. Each one of you will have ten tokens at the beginning of each round, and you will play this game with each one of your classmates once.
The name of the classmate you will be playing with will be shown in the screen in each round.

\begin{figure}[htb!]
\centering
\includegraphics[width=0.7\linewidth]{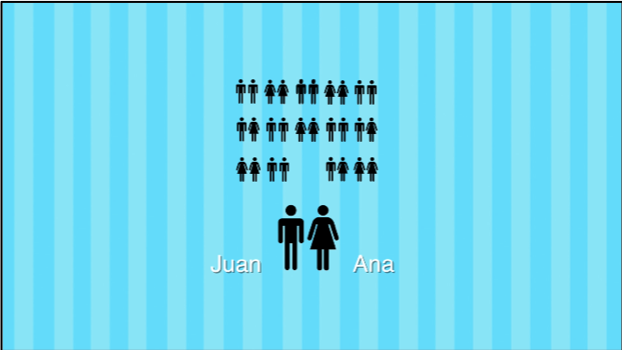}
\label{SM_protocol_1}
\end{figure}

In your tablet you will see a circle in the center of the screen and your ten tokens in the right side of it. When everybody is ready to play you will be able to move the tokens you want to the center. Each token that you move to the center will be multiplied by two and these tokens will be sent to your classmate. If you decide to move one token to the center, you will see that there will appear two tokens. If you move two tokens there will appear four tokens and so on.

\begin{figure}[htb!]
\centering
\includegraphics[width=0.7\linewidth]{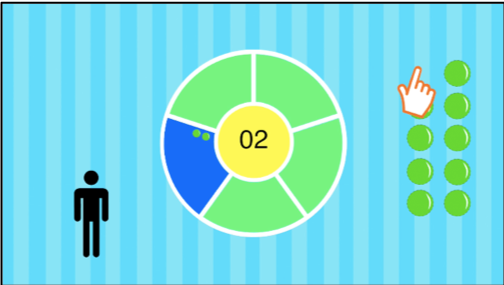}
\label{SM_protocol_2}
\end{figure}

Same thing will be going on in your classmates’ screen.
Everything you must do is to decide how many tokens you want to cooperate with your partner and how many you want to keep. If you are ready with your decision, you must click send.  When everybody had already played, the game will assign you a new classmate, and the next round will start. You will have again ten tokens to play, no matter what happened in the previous rounds.

\begin{figure}[htb!]
\centering
\includegraphics[width=0.7\linewidth]{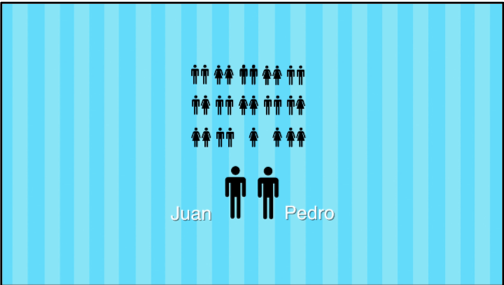} 
\label{SM_protocol_3}
\end{figure}

At the end of the n-th round, everyone will have had interacted once with each other classmates and the game will be over. Then, you will be informed how many tokens you will have obtained after the game.

\begin{figure}[htb!]
\centering
\includegraphics[width=0.7\linewidth]{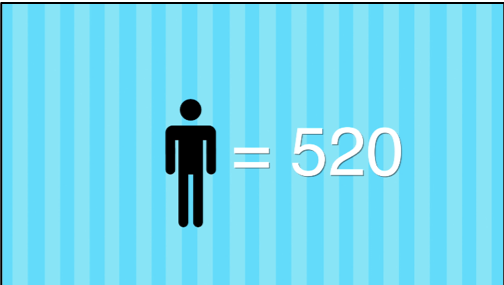}
\label{SM_protocol_4}
\end{figure}

Do you see? It is very simple. Let’s review one more time.
You have ten tokens at the beginning of each round.
You will see the classmate you will play with and you will play with each one of your classmates.
You decide how many tokens you will give to your partner, which will multiply by two and will be kept by your classmate.
The number of tokens you receive will multiply by two and will be added to the tokens you kept.
The game will be over when you have played with all your classmates. And then you will know how many tokens you earned at the end of the game.
Thanks for your participation!”

Stage 5: Closure of the activity
When all the students have finished, fieldwork team collects the tablets and headphones.
Once tablets are collected, fieldwork attendant says:
“Congrats, you made a great work. Did you like the game?” [Wait for reactions]. 
If nobody talks, make some open questions about the experience.
After thanking the teacher, fieldwork team leaves the classroom.

\subsubsection*{FAQ}
Each question contains the explanation for teachers or implementers, and a possible version of answer for children shown in \textit{italics}. 

\begin{itemize}
 \item What happen if there are absent students?
 At the beginning all students of the class list are registered on tablets. Nevertheless in the moment of implementation the teacher or monitor will choose only the present students to play the games. For surveys, possible answers of absent students are considered.
 
 \item What happen if a student withdraws during the activity?
 It is important to attempt this not happen. Nevertheless, if a student needs to leave the classroom the teacher or monitor will take his or her place and will let the children know it. The adult who replaces the student must always play giving the half of their tokens.
Answer for children: \textit{Peter has to leave because of X, so in order to we can continue the game, me, Mr. Z will take his place and I will always give 5 tokens.}

\item How is the winner determined? Who wins? How many tokens you need to earn to win?
In this type of games nobody wins or loses. Depending on the actions of each student he or she will finish the game with a total number of tokens but the game is designed so no student can visualize clearly the level of cooperation he or she received in each round or interaction with each classmate.
A standard answer to this would be: \textit{The game is made so nobody wins or loses, but everybody gets fun while each one piles up different amounts of tokens.}

\item How many tokens do I have? Do I lose them from one round to the other? 
In each round the students have ten tokens no matter what happened in the previous rounds. They never lose their tokens that they keep or receive because all of them are added to the final amount.
Possible answer: \textit{You will always start the round with 10 tokens and you won’t lose any that you have kept or received from another partner.}

\item Can I go back and play again with a classmate who I have already played with?

In game 1 the groups are formed randomly in each round. It is possible that you play more than one time with the same classmate and in some occasions you won’t play with a specific classmate in this game.
In game 2 you will always know who you are playing with, but you will play one time with each classmate in the room so you won’t be able to go back and play again with a specific classmate.
Possible answer: \textit{No, because the idea is that each one plays with the maximum of possible partners (game 1); No, because you are going to play once with each of your classmates and if we repeat a classmate we wouldn’t have time enough to play with everyone (game 2).}

\item Can I hoard tokens from one round to another?

The rounds are programed to be independent from each other because information that each one of them picks up is different, looking for information about the level of cooperation with each one of the classmates in the same situation. Otherwise, the order in which classmates are paired would influence on every student path and endownment, and then we could not compare interactions between students without taking into account the chronological order and history in the game.
Possible answer: \textit{No, when you begin each round of each game you start with ten tokens no matter the outcome of previous rounds.}

\item How much time do I have to decide my move?

There is no limit of time; nevertheless if students take too much time the game will be too slow for everybody. Because of that implementers must help belated children basing on the information shown in control screen.
Possible answer: \textit{There is no limit of time, but the idea is that you play as fast as possible so that it will be funnier for everybody. If you have doubts or any problem please let me know and I will help you.} 
 
\item My tablet does not sound / It turned off / I do not have Internet signal /Etc.
In these situations implementers must check the Tablet / Correct if possible / Re assign a new tablet if it happens in the beginning of the game.
Possible answer: \textit{Let me see the problem. Don’t worry; we will solve it quickly.} 

\item Can I choose who I am going to play with in the next round?
The program selects randomly the game partner until make sure that each student played with all of his or her classmates. 
Possible answer: \textit{No, because the programs makes couples randomly. } 
 
\item How can I know how many tokens I have before finishing the game?
The feedback of how many tokens each student receives in the pit game is shown immediately after finishing the round. This is because in this type of games it is important to know the outcome of each action and in the pit game there is “common pit” cooperation so no matter the result any child will feel affected.

In the couple game, it is crucial to prevent that no child feels hurt or affected because of the number of tokens that he or she received from each classmate, and we also don't want to provoke later consequences between relations, and to affect student's evaluation of the next classmate because of previous history with others in previous rounds. Because of that, the outcome feedback is added into a total amount that is shown in the end of the game.
Possible answer: \textit{You need to finish the game with all of your classmates and then at the end you will know the total amount of tokens you got.}

\item Can I exchange tablets with my desk classmate? Can I ask for help or help a classmate? Can I say my partner how much he or she gave me or what I gave him or her?
Both the surveys and the games are an individual exercise. It is important to prevent any conversation between classmates during the implementation and therefore it must be checked that all students have their headphones on.

Possible answers: \textit{No, you have to answer by your own, with your assigned tablet, and you can not share or talk with your classmates during the game because otherwise the game would not be the same. If you have any doubt I will be here to help you. } 

\item Must I respond all questions in the survey?
It is very valuable to have all information that can be collected if all questions in survey are responded, but if there is a question that students cannot or do not want to answer they must not be pressured to do it.
Possible answer: \textit{Try to answer all questions that you can, if you have any doubt you can ask me. If there is any question that you can’t or don’t want to answer, don’t worry, you can leave it blank.} 

\end{itemize}

\pagebreak

\subsection*{SM 2.3: Survey Description}

Peer Nomination Survey

Students were asked to nominate their classmates on a variety of social and personality characteristics and friendship. An alphabetically sorted roster with each other classroom student name was displayed for each of the nine questions, and they were told to nominate at least two and a maximum of five peers per question, without any other restriction. Questions were the following (in Spanish and then a translation):

\begin{enumerate}
    \item Selecciona \textbf{2, 3, 4 o  5 compañeros/as} con los cuales \textbf{M\'AS TE GUSTARÍA JUNTARTE.}
    \item Selecciona \textbf{2, 3, 4 o  5 compañeros/as} con los cuales \textbf{MENOS TE GUSTARÍA JUNTARTE.}
    \item Selecciona \textbf{2, 3, 4 o  5 compañeros/as} a quienes consideras \textbf{TUS MEJORES AMIGOS.}
    \item Selecciona \textbf{2, 3, 4 o  5 compañeros/as} a quienes consideras que son  \textbf{LOS MÁS POPULARES Y CONOCIDOS} en tu curso.
    \item Selecciona \textbf{2, 3, 4 o  5 compañeros/as} a quienes consideras que  \textbf{NO SON POPULARES NI CONOCIDOS} en tu curso.
    \item Selecciona \textbf{2, 3, 4 o  5 compañeros/as} a quienes consideras que tienden a  \textbf{INICIAR PELEAS.} Empujan a otros, o les pegan o patean.
    \item Selecciona \textbf{2, 3, 4 o  5 compañeros/as} a quienes consideras que  \textbf{SE RÍEN DE LA GENTE.} Les gusta burlarse de otros y avergonzarlos en frente de otras personas.
    \item Selecciona \textbf{2, 3, 4 o  5 compañeros/as} as a quienes consideras que  siempre están dispuestos a hacer algo  \textbf{AMABLE PARA OTRAS PERSONAS,} o son personas muy amistosas.
    \item Selecciona \textbf{2, 3, 4 o  5 compañeros/as} a quienes consideras que \textbf{ COOPERAN.} Siempre ayudan, comparten y dan un turno a todos.
    
\end{enumerate}

Construction of variables:
All received nominations were counted for each student. We used the following constructs:

Perceived popularity: We calculate the number of nominations in question 4 (most popular) minus the number of nominations in question 5 (not popular/unpopular).

Aggressiveness: We calculate nominations in question 6 plus nominations in question 7.

Prosociality: We calculate nominations in question 8 plus nominations in question 9.

Liked: We calculate nominations in question 1
Disliked: We calculate nominations in question 2

Friendship Nominations: We calculate nominations in question 3

Then, we standardize all these coefficients, subtracting the classroom mean and dividing by their classroom standard deviation.

\pagebreak

\section*{Supplementary Figures}
\subsection*{SM 3.1: Sent tokens by students}

\begin{figure}[!htb]
\centering
\includegraphics[width=0.9\linewidth]{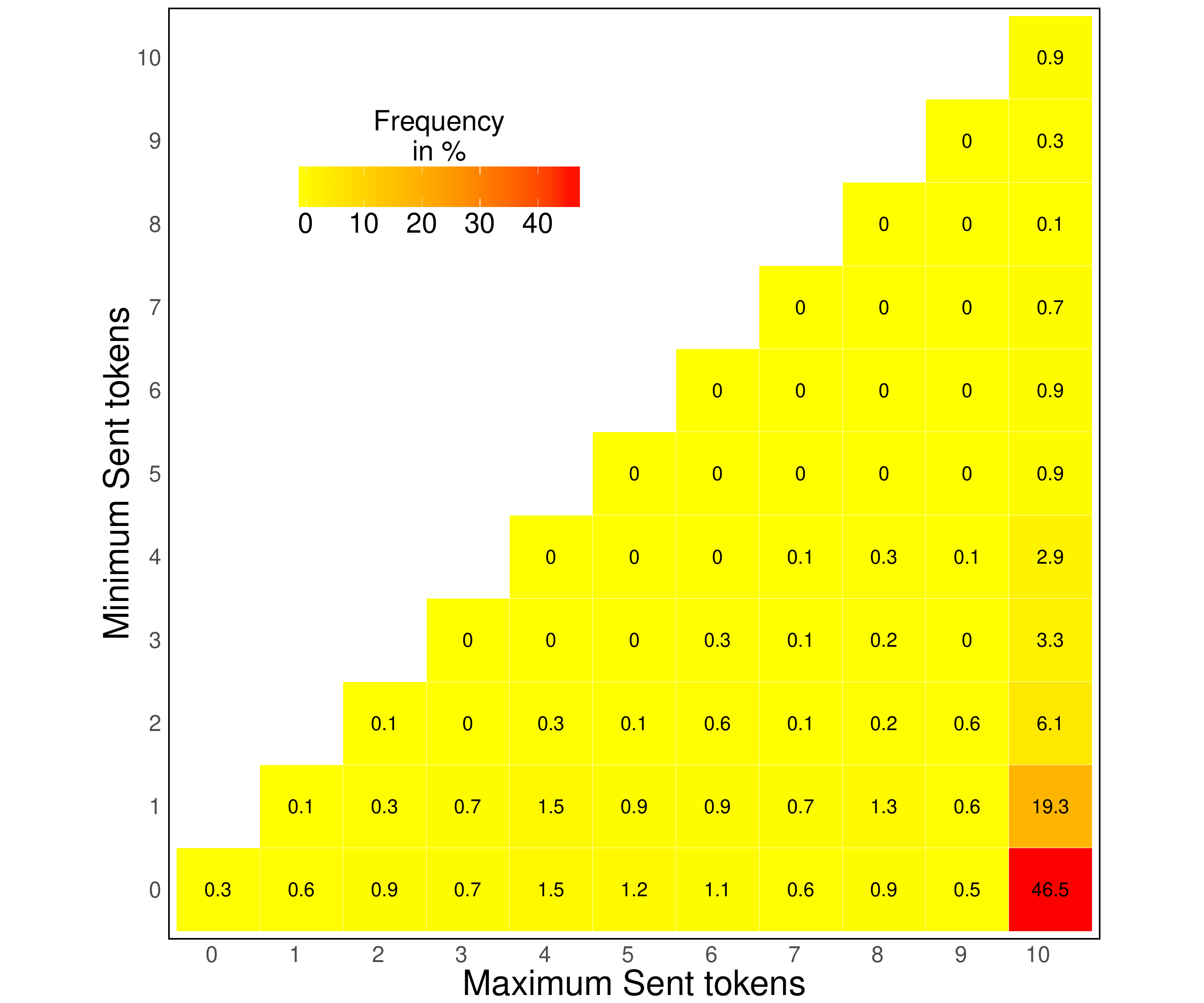}
\caption{Heatmap for maximum and minimum sending. At the bottom-right corner, we observe a cluster ($ \approx 65\%$ of students, red/orange blocks) indicating that students send tokens in the whole range of possibilities. History between colleagues matters.}
\label{SM_heat}
\end{figure}

\pagebreak

\subsection*{SM 3.2: Correlation Matrix}

\begin{figure}[!ht]
\centering
\includegraphics[width=0.95\linewidth]{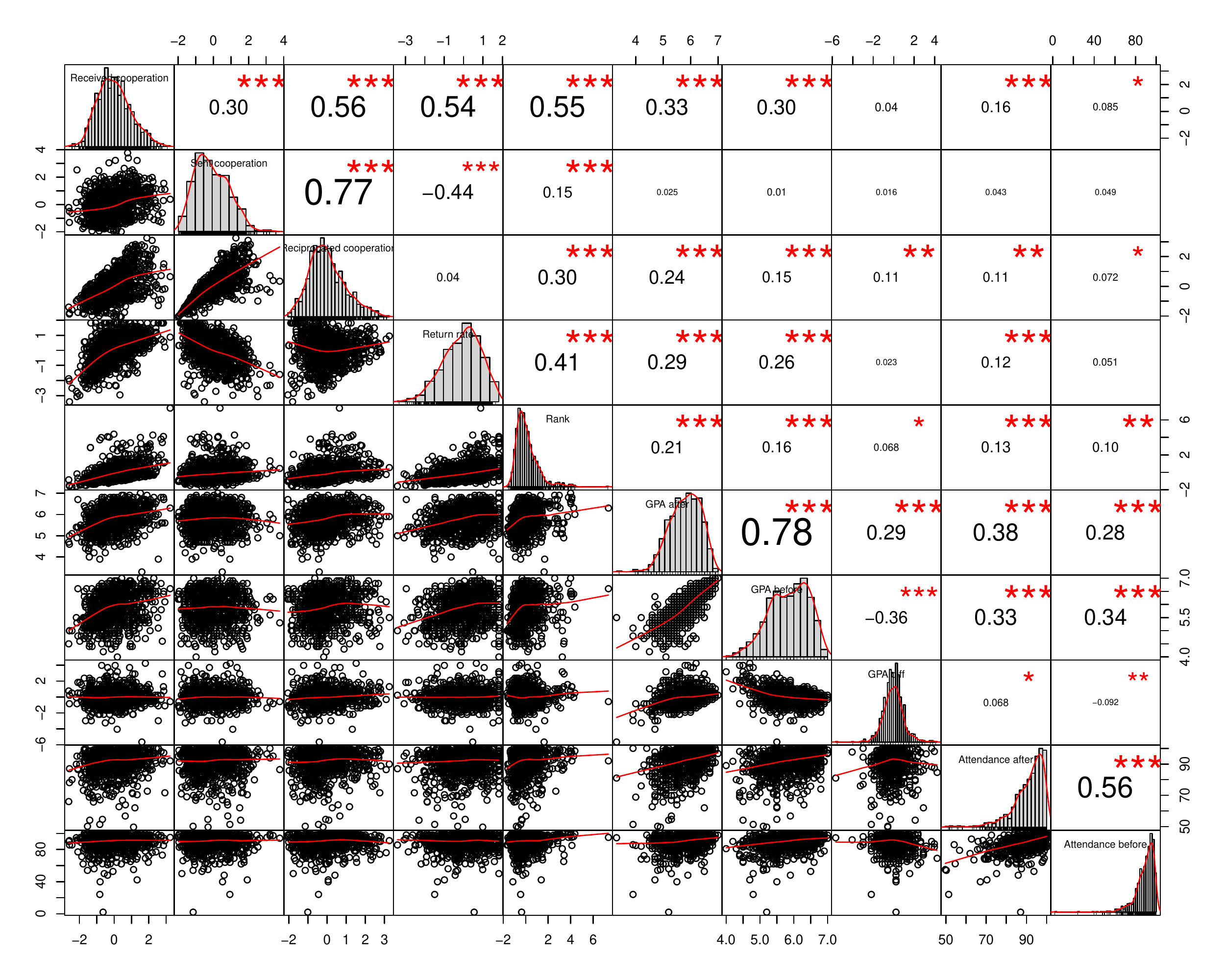}
\caption{Correlations between all collected, measured, and built variables.}
\label{correlation_matrix}
\end{figure}

\pagebreak

\subsection*{SM 3.3: Interaction between friendship and GPA}

\begin{figure}[htb!]
\centering
\includegraphics[width=0.9\linewidth]{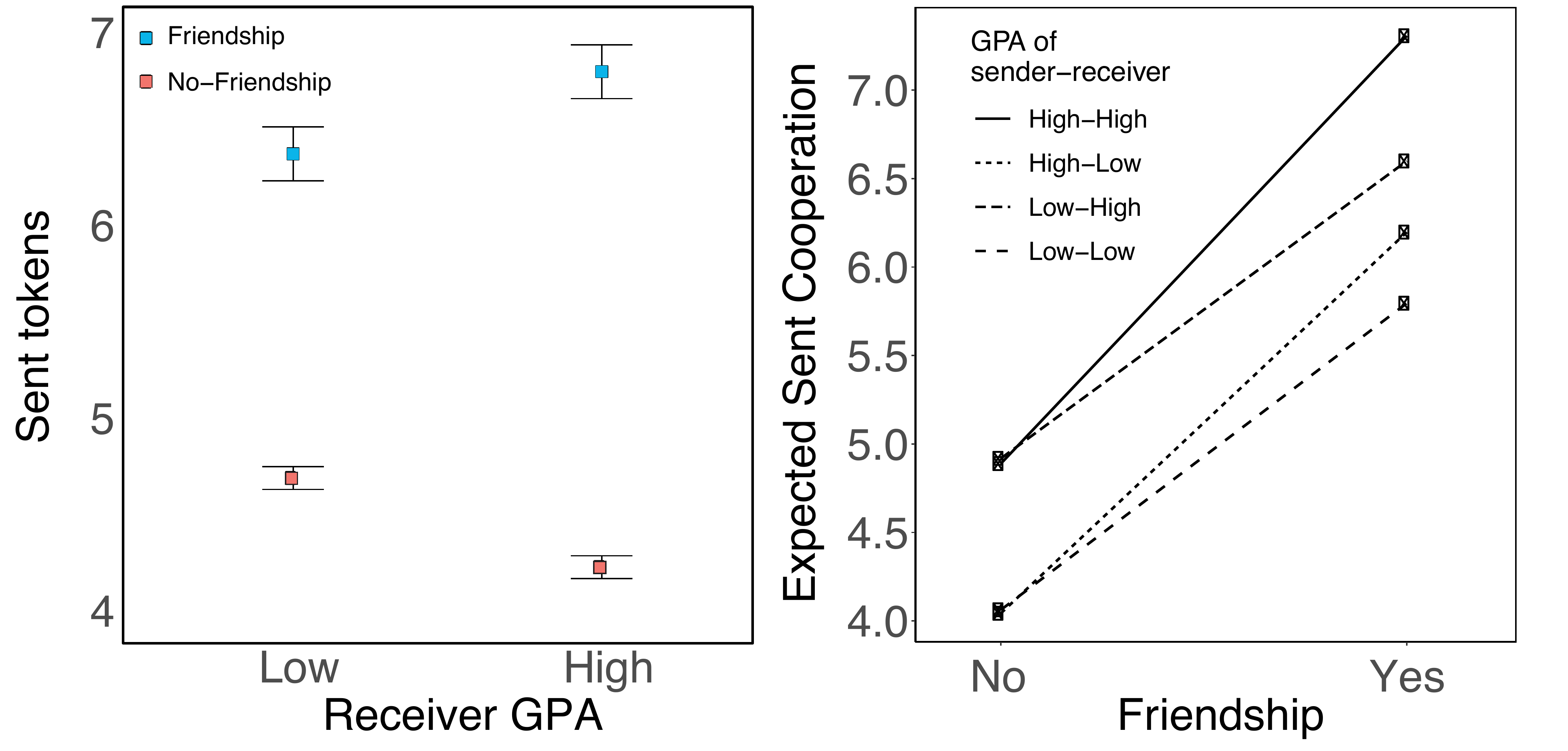}
\caption{A) Sent cooperation. Red represents non-friendship relationships and cyan represents friendship relationships. We observe that high GPA students sent more cooperation to their friends than low GPA students. B) Interaction effect between friendship and the GPAs of the sender and receiver}
\label{friend_eff}
\end{figure}

Figure \ref{friend_eff} B shows the interaction effect between friendship and the GPAs of the sender and receiver in a dyad on the number of sent tokens. The interaction effects are significant (p-value $<0.01$) and positive, meaning that the slopes associated with friendship are larger for high GPA than low GPA students. This means that high GPA students are more strategic in how they invest their cooperation, investing more in friendships, specially, with other high GPA students. We also observe that high GPA students receive more cooperation overall, suggesting that on average students understand that investing on high GPA peers is strategic, but high GPA students place more emphasis on this strategy.

\pagebreak

\subsection*{SM 3.4: Cooperative strategies of students}

\begin{figure}[!htb]
\centering
\includegraphics[width=0.99\linewidth]{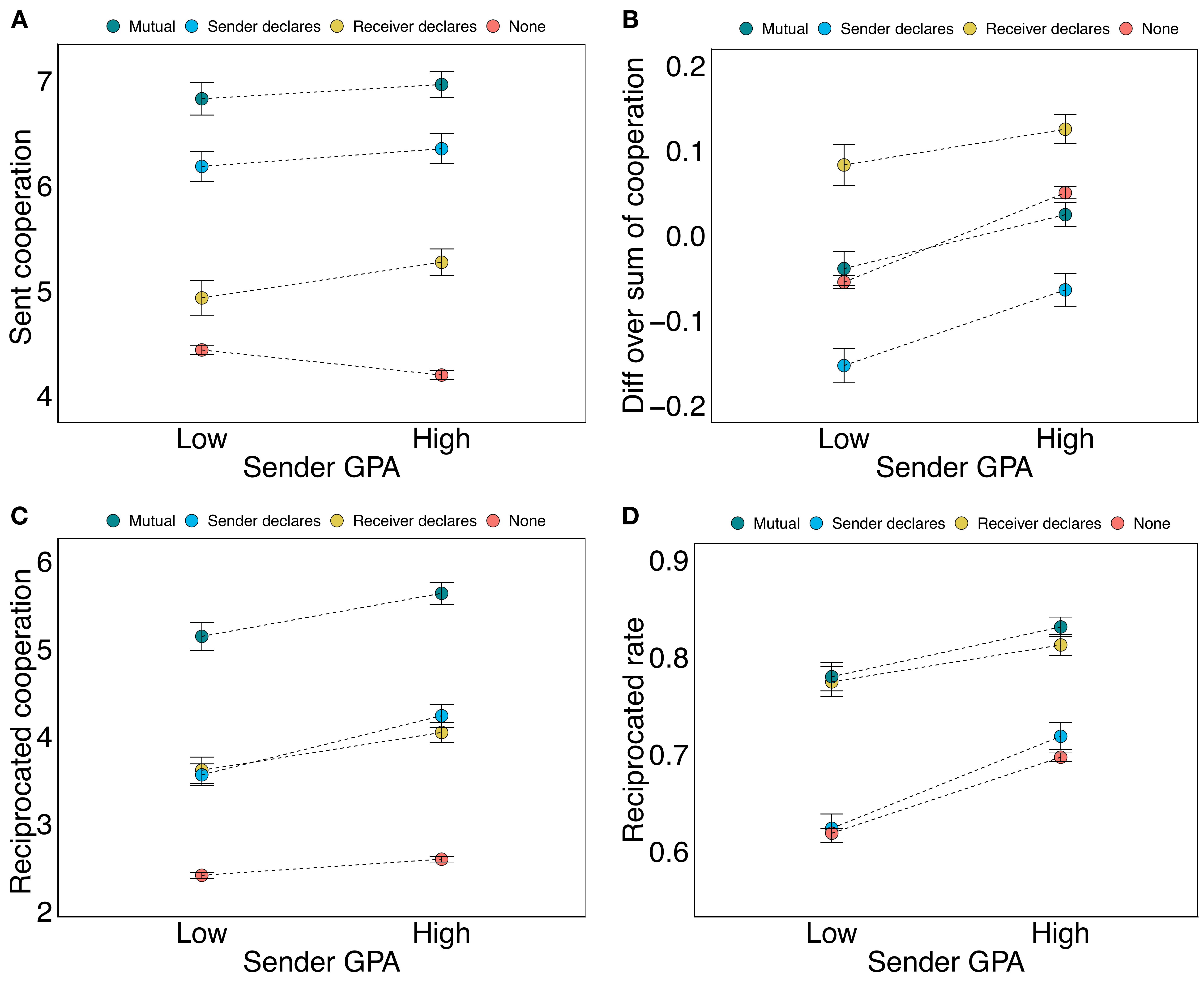}
\caption{Cooperative strategies of elementary school children, considering declared friendship (vertical lines) and the GPA of the sender (x-axis). A) Sent cooperation. B) Ratio of the received (s) minus sent (r) cooperation over received plus sent cooperation ($(r_i-s_i)/(r_i+s_i)$). C) Reciprocated cooperation. D) Reciprocated rate (reciprocated cooperation over sent cooperation).}
\label{desc_11}
\end{figure}

\begin{figure}[!htb]
\centering
\includegraphics[width=0.99\linewidth]{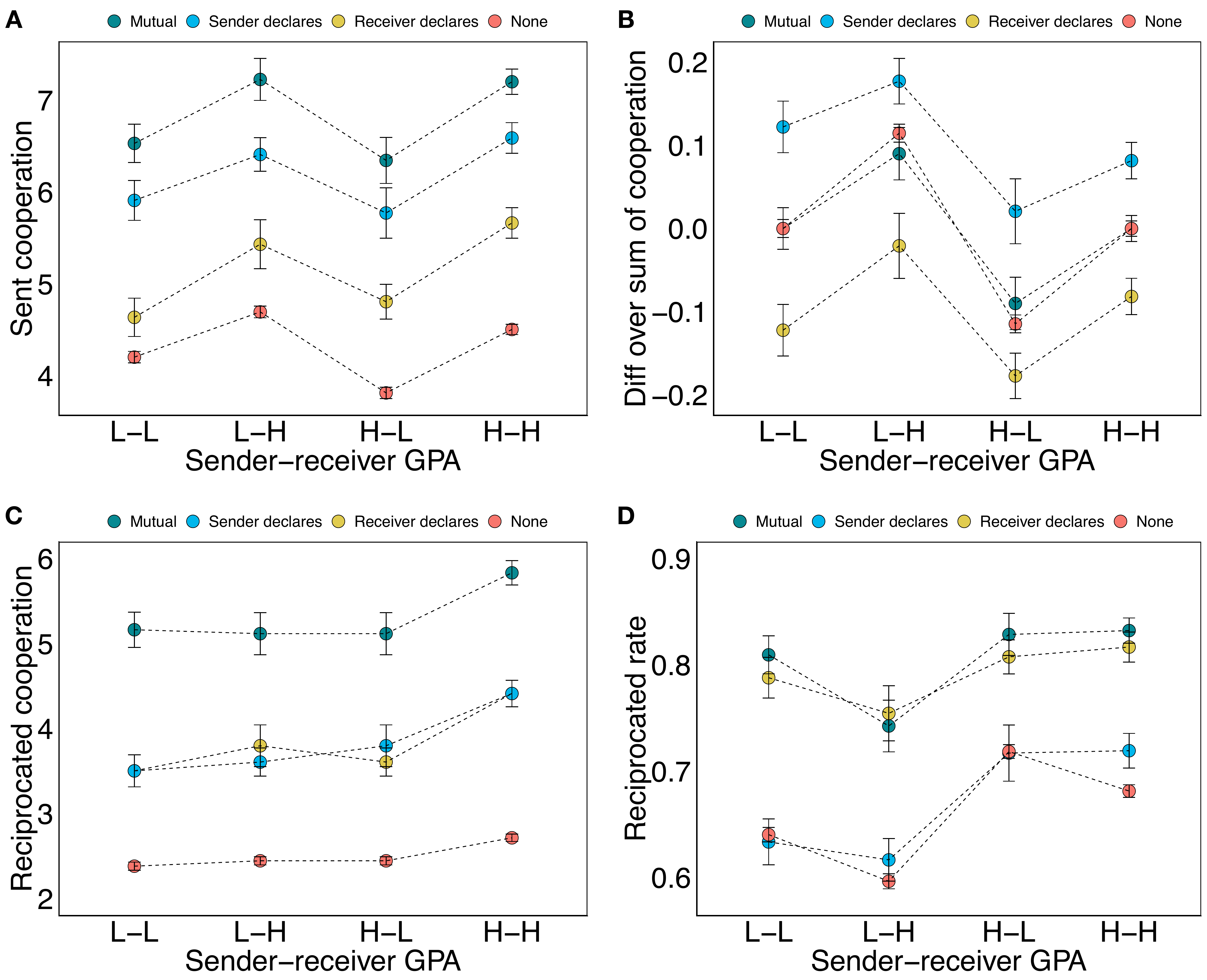}
\caption{Cooperative strategies of kids in elementary schools, considering declared friendship (vertical lines) and GPA of sender and receiver (x-axis). A) Sent cooperation. B) Ratio of the received (s) minus sent (r) cooperation over received plus sent cooperation ($(r-s)/(r+s)$). C) Reciprocated cooperation. D) Reciprocated rate (reciprocated cooperation over sent cooperation).}
\label{strat_1}
\end{figure}

\clearpage

\subsection*{SM 3.5: Explanatory Power of the Model}

\begin{figure}[!htb]
\centering
\includegraphics[width=0.6\linewidth]{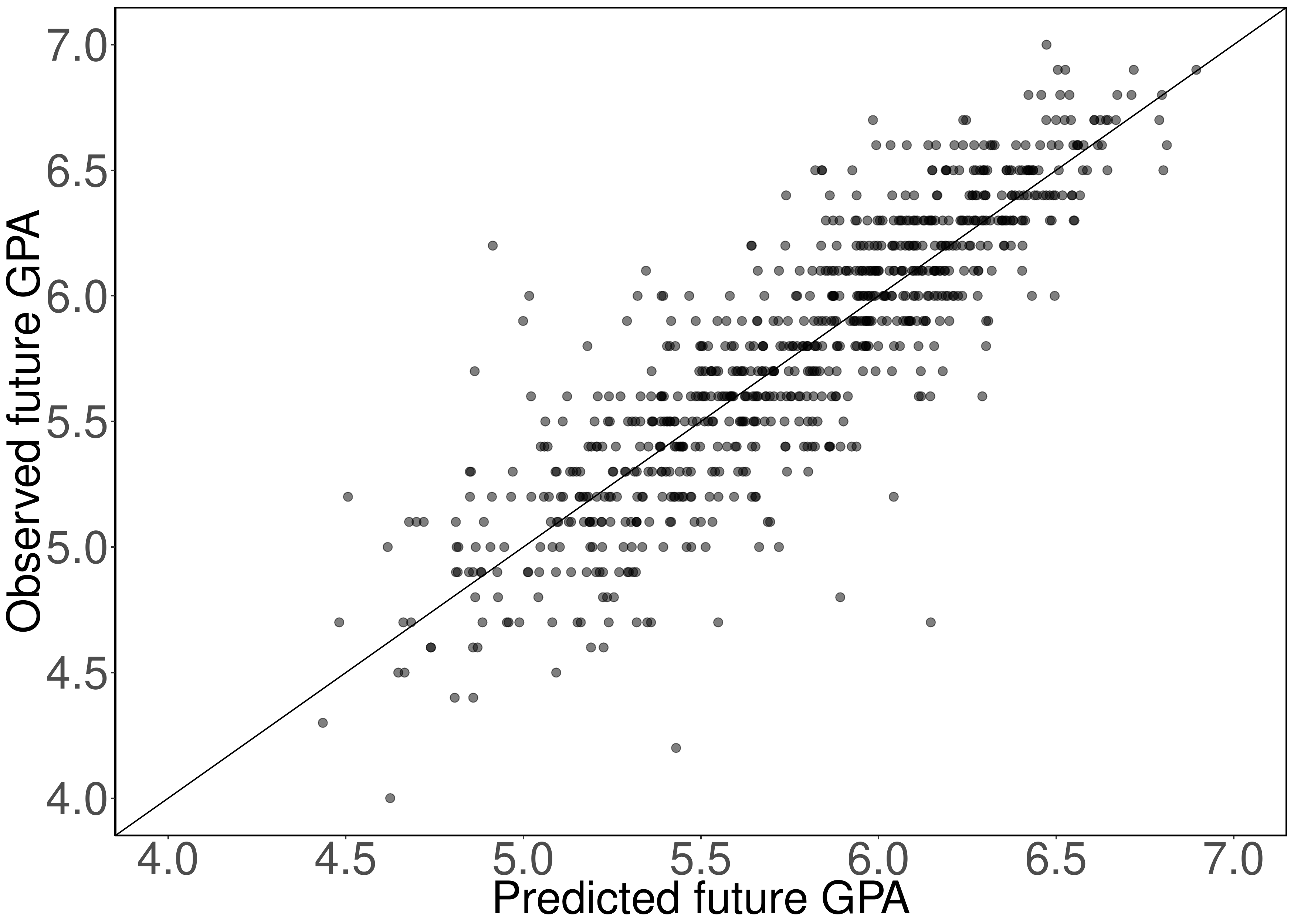}
\caption{Observed v/s predicted GPA (table 2 model 4)}
\label{adj_regg1}
\end{figure}

\begin{figure}[!ht]
\centering
\includegraphics[width=0.6\linewidth]{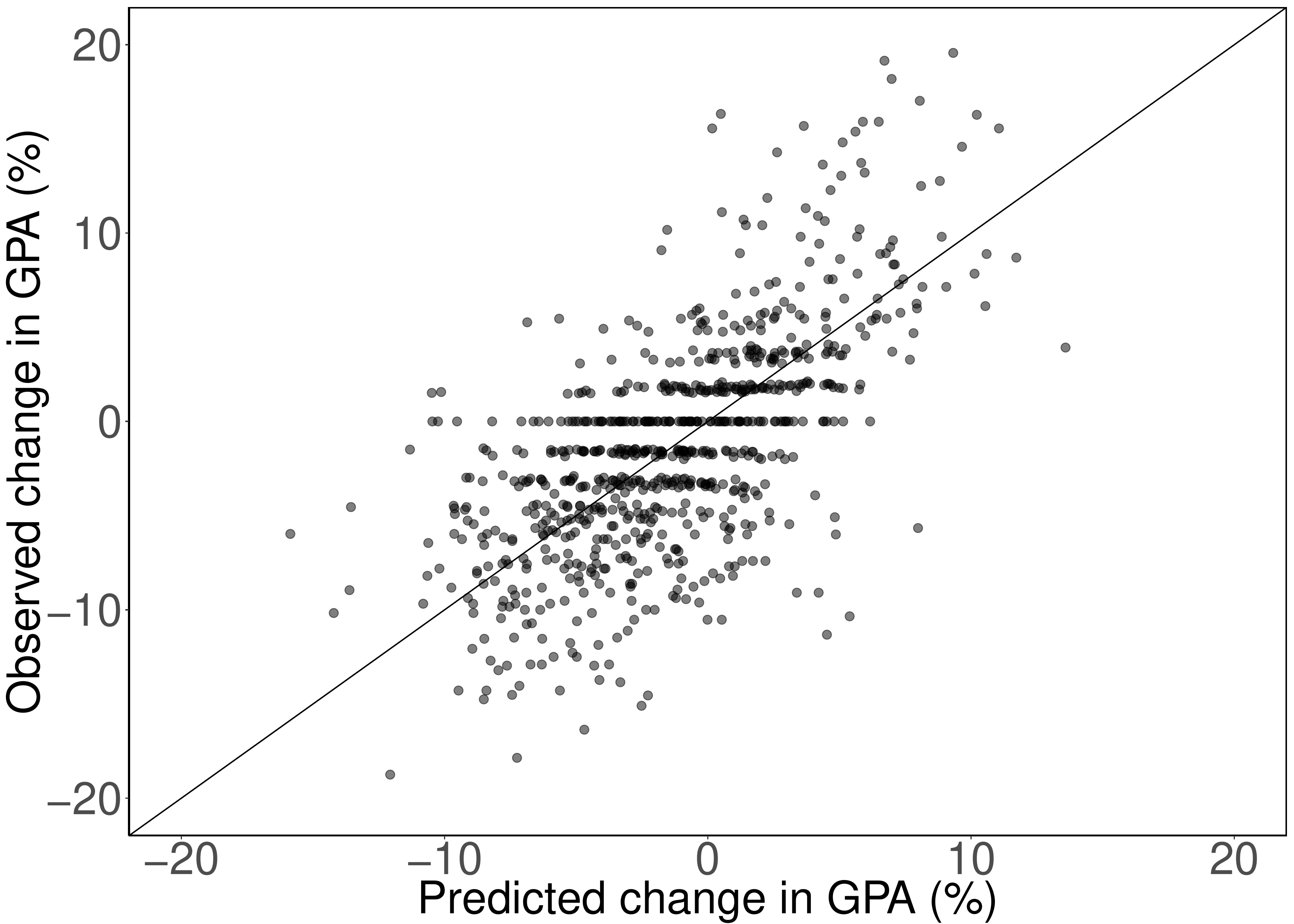}
\caption{Observed v/s predicted change in GPA .}
\label{adj_regg2}
\end{figure}

\pagebreak

\subsection*{SM 3.6: Friendship Relationships}

\begin{figure}[!ht]
\centering
\includegraphics[width=0.95\linewidth]{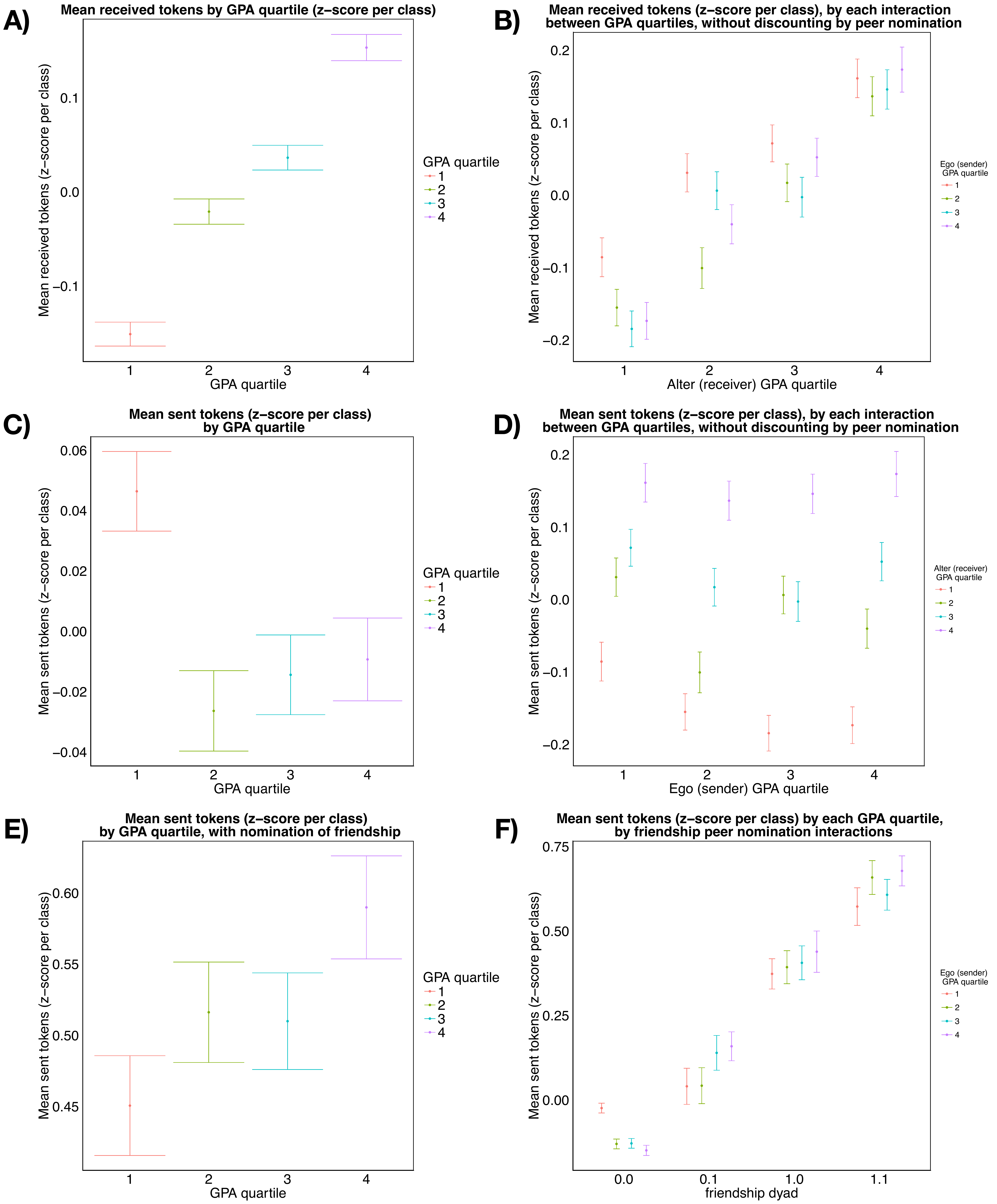}
\caption{A) Relationship between received tokens and GPA quantile of the receiver. B) Decomposition of figure (A) by GPA quantile of the sender. C) Relationship between sent tokens and GPA quantile of the sender. D) Decomposition of figure (C) by the GPA quantile of the receiver. E) Relationship between sent tokens and GPA quantile of the sender considering just friendship peer-nominations. F) Decomposition of figure (E) by the type of friendship relationship.}
\label{firend_sent2}
\end{figure}

\pagebreak

\subsection*{SM 3.7: Sent and received tokens by GPA levels, mediated by other peer-nomination sociobehavioral characteristics }

\begin{figure}[!ht]
\centering

\includegraphics[width=0.95\linewidth,height=0.78\textheight,keepaspectratio]{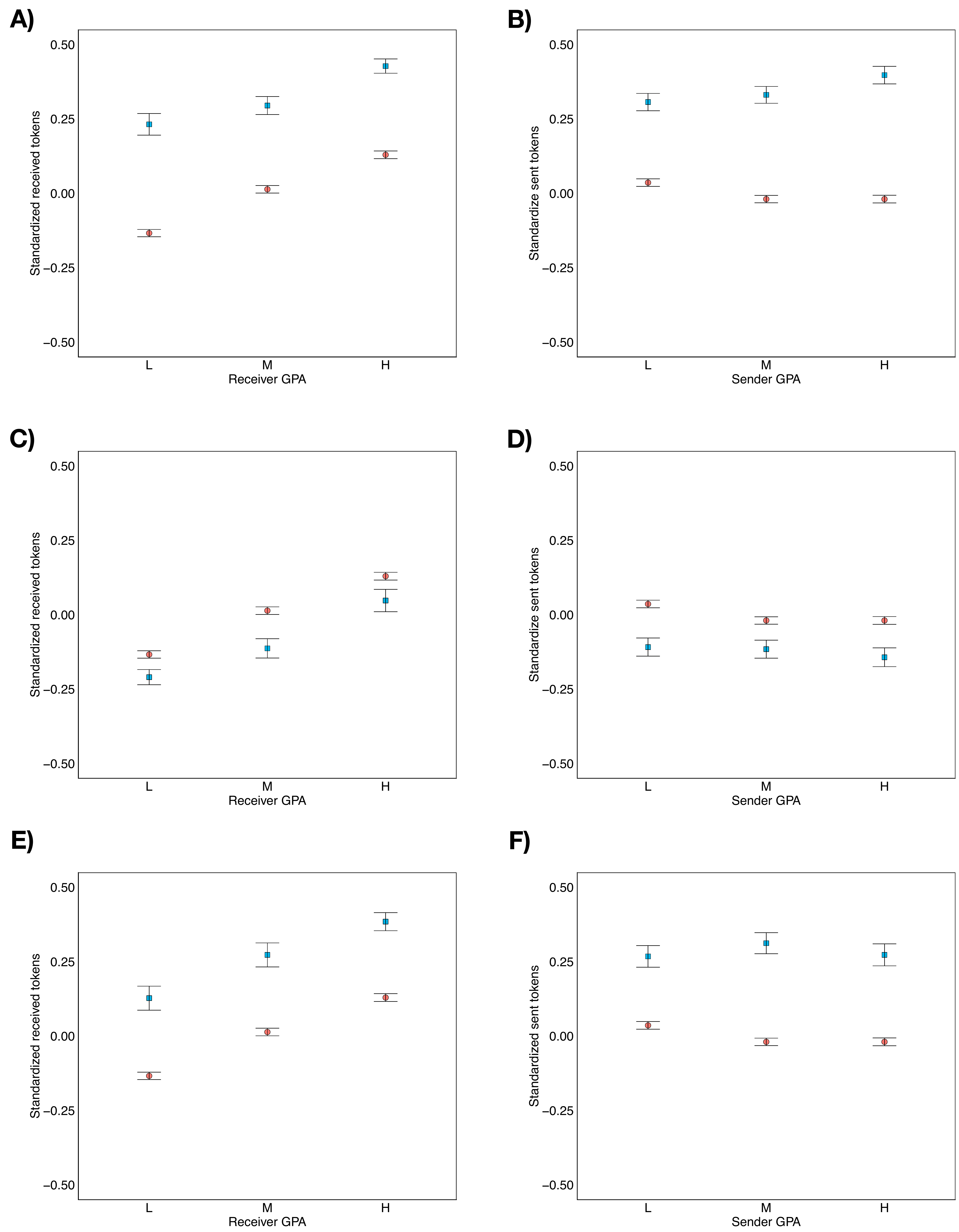}
\caption{{A) Relationship between received tokens and GPA level of the receiver, comparing nominated prosocial student dyad  v/s all dyads B) Relationship between sent tokens and GPA level of the receiver, comparing nominated prosocial student dyad v/s all dyads C) Relationship between received tokens and GPA level of the receiver, comparing nominated aggressive student dyad v/s all dyads D) Relationship between sent tokens and GPA level of the receiver, comparing nominated aggressive student dyad v/s all dyads E) Relationship between received tokens and GPA level of the receiver, comparing nominated popular student dyad v/s all dyads F) Relationship between sent tokens and GPA level of the receiver, comparing nominated popular student dyad v/s all dyads. All dyads are in red, nominations are in cyan.}}
\label{SM_panels}
\end{figure}

\end{document}